\documentclass[aps,prl,twocolumn,showpacs,reprint,superscriptaddress]{revtex4-1}

\usepackage{amsmath, amsfonts, amssymb, bm}
\usepackage{float}
\usepackage{mathtools}
\usepackage[inline]{enumitem}
\usepackage{graphicx}
\usepackage{lineno}
\usepackage{xcolor}
\usepackage[breaklinks,hypertexnames=false]{hyperref}
\hypersetup{colorlinks,linkcolor={magenta},citecolor={blue},urlcolor={blue}}
\usepackage[capitalize]{cleveref}

\newcommand{\clr}[1]{\textcolor{blue}{#1}}

\begin{document}
\title{Tunable Josephson diode effect on the surface of topological insulators}

\newcommand{\tianjin}{Department of Physics, Tianjin 300072, China}
\newcommand{\tjlab}{Tianjin Key Laboratory of Low Dimensional Materials Physics and Preparing Technology, Tianjin 300072, China}

\newcommand{\uam}{Department of Theoretical Condensed Matter Physics, Condensed Matter Physics Center (IFIMAC) and Instituto Nicol\'as Cabrera, Universidad Aut\'onoma de Madrid, 28049 Madrid, Spain}

\newcommand{\nagoya}{Department of Applied Physics,
	Nagoya University, Nagoya 464-8603, Japan}

\newcommand{\RIKEN}{Center for Emergent Matter Science (CEMS), RIKEN, Wako, Saitama 351-0198, Japan}
\newcommand{\tokyo}{Department of Applied Physics, The University of Tokyo, Tokyo 113-8656, Japan}

\author{Bo Lu}
\affiliation{\tianjin}

\author{Satoshi Ikegaya}
\affiliation{\nagoya}

\author{Pablo Burset}
\affiliation{\uam}

\author{Yukio Tanaka}
\affiliation{\nagoya}

\author{Naoto Nagaosa}
\affiliation{\RIKEN}
\affiliation{\tokyo}

\date{\today}

\begin{abstract}
The Josephson rectification effect, where the resistance is finite in one direction while zero in the other, has been recently realized experimentally. 
The resulting Josephson diode has many potential applications on superconducting devices, including quantum computers. 
Here, we theoretically show that a superconductor-normal metal-superconductor Josephson junction diode on the two-dimensional surface of a topological insulator has large tunability. 
The magnitude and sign of the diode quality factor strongly depend on the external magnetic field, gate voltage, and the length of the junction. 
Such rich properties stem from the interplay between different current-phase relations for the multiple transverse transport channels, and can be used for designing realistic superconducting diode devices. 
\end{abstract}
\maketitle

\emph{Introduction.---}  
The diode exhibits a large difference in resistance between opposite bias polarities, making it one of the fundamental electronic components. 
This nonreciprocal effect requires inversion symmetry breaking, which is usually achieved in heterostructures like the pn-junction, where the current-voltage (I-V) characteristic is fixed once the structure is fabricated (except for its temperature dependence). Consequently, the direction of the rectification cannot be reversed. 
Similar nonreciprocal transport properties have been studied in noncentrosymmetric bulk materials, like, e.g., magneto-chiral anisotropy~\cite{Rikken1997,Rikken01,Rikken02,Rikken05,Pop2014,Morimoto16}, but the rectification effect is typically less than $1\%$. 
%
%
Recently, the superconducting diode effect (SDE), or Josephson diode effect (JDE) when implemented on Josephson junctions, has excited great fundamental and applications-oriented interest. The SDE allows for a larger critical supercurrent in one direction than in the opposite. This nonreciprocity can be expressed by the quality factor $Q$ in the form of forward/reverse critical currents $J_c^{\pm}$,
\begin{equation}
	Q=\frac{J_{c}^{+}- J_{c}^{-} }{J_{c}^{+}+J_{c}^{-} }.
\end{equation}
As $Q\neq 0$, the system admits a perfect dissipationless transmission in only one direction if currents flow in the range between $J_{c}^{+}$ and $
J_{c}^{-} $. Such a directionality provides a promising avenue for future applications to diode devices, such as dissipationless electronics. 
SDE has been revealed in various systems such as noncentrosymmetric superconductors~\cite{OnoNat,Miyasaka_2021,zhang2020,Wakatsuki17,Hoshino18}, patterned superconductors~\cite{lyu_2021,bauriedl_2022}, superconductor-ferromagnet multilayers~\cite{narita_2022,strambini_2022,Golubov22}, and twisted materials~\cite{lin_2022,HuXin,Alvarado23}. In particular, a lot of experimental efforts have been devoted to the JDE~\cite{Bocquillon16,Baumgartner22,Wu22,Merida21,Pal22,Baumgartner_2022,Gupta22,Turini22}, where a large quality factor $Q$ is expected.

\begin{figure}[b]
	\begin{center}
		\includegraphics[width=80mm]{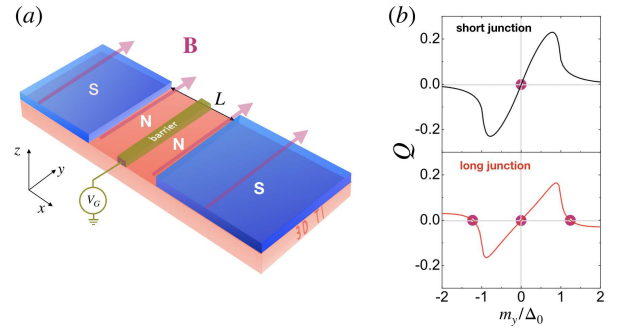}
	\end{center}
	\caption{(a) Schematic of a planar SNS Josephson junction formed on the surface of a 3D TI, with a normal region of length $L$ and an in-plane magnetic field $B$ applied to the whole structure. Electrically tunable gate $V_G$ can be applied on the thin barrier in the middle. (b) Characteristic of the quality factor $Q$ in short (top) and long (bottom) junctions. $Q$ in the long junction experiences several sign changes (purple circles) as the Zeeman field varies. }
	\label{Fig1}
\end{figure}

Most theories of the JDE~\cite{Jiangping07,Zhang21,Misaki21,Kopasov21,He_2022,Yanase22,Yuan22,Bergeret22,Halterman22,Karabassov22,Souto22,Sigrist22,Fominov22,Davydova22,Tanaka22,Wang2022} predict that the absence of inversion and time-reversal symmetries results in the anomalous Josephson current $J\left( \varphi \right) \neq -J\left( -\varphi \right) $, with $\varphi $ being the macroscopic superconducting phase difference. 
Consequently, Cooper pairs acquire a finite momentum $q_{0}$, like Fulde-Ferrell-Larkin-Ovchinnikov states, resulting in different depairing effects between supercurrents with opposite directions~\cite{Yuan22}. 
Since $q_{0}$ can be induced by magnetic fields in spin-orbit coupling systems, topological materials have advantages in revealing JDE due to their strong spin-orbit coupling. JDEs have been thus predicted in one-dimensional (1D) nanowire systems~\cite{Reynoso12}, quantum spin Hall insulators~\cite{Meyer15}, and the surface of topological insulators (TIs)~\cite{Alidoust17,Tanaka22}. 

Notably, three issues remain unexplored among previous works on the JDE in topological materials. 
First, JDEs have been studied in short junctions, while most experiments so far are based on long Josephson junctions. Generally, the length of the junction increases the number of resonances inside the normal region and thus the behavior of the Josephson current~\cite{Kulik,Ishii70,Bardeen72}. 
Second, most microscopic models considered effective one-dimensional transport, while the JDE in actual higher dimensional systems is more relevant for applications since it supports large supercurrent flow. Moreover, higher dimensional junctions consist of multiple channels which may greatly alter the JDE, specially in the presence of chiral Majorana bound states~\cite{Tanaka22}. It is thus worth studying the JDE beyond short and one-dimensional junction limit. 
The last remaining issue is how to easily tune the JDE by external fields or junction parameters with respect to both the sign and magnitude of the quality factor. 

In this Letter, we study the Josephson diode effect in a superconductor-normal metal-superconductor (SNS) junction on the surface of a three-dimensional (3D) TI~\cite{Fu07,Zhang09,Hsieh08,Hsieh09,xia_2009,Chen09,Hasan09,Ren10}, as depicted in Fig. \ref{Fig1}(a). 
A tunable gate voltage $V_G$ can be applied in the thin barrier region in the middle of the junction. In short junctions, $Q$ has no sign change unless the direction of the applied in-plane Zeeman field $m_{y}$ flips, see Fig. \ref{Fig1}(b). However, $Q$ can exhibit sign changes as a function of the magnitude of $m_{y}$ in the long junction without voltage $V_G$. Furthermore, a finite $V_G$ can be used to tune the quality factor $Q$. 
To explain this behavior, we show how the interplay between the magnetic field and the spin-orbit coupling on the TI surface results in a different momentum shift $q_{0}$ for each transport channel. 
Consequently, since the supercurrent is carried by multiple modes with different momenta parallel to the interface, the current-phase relations can be different among channels, like in the $d$-wave junctions~\cite{Tanaka97}. 
The total supercurrent and the quality factor $Q$ can thus be greatly modified when the current-phase relation for different transverse modes deviates from each other by varying parameters, such as the gate voltage. We note that our computed $Q$ is consistent with recent experiments~\cite{Pal22} and useful for designing realistic rectification devices based on TIs. 

\begin{figure}[tb]
	\begin{center}
		\includegraphics[width=85mm]{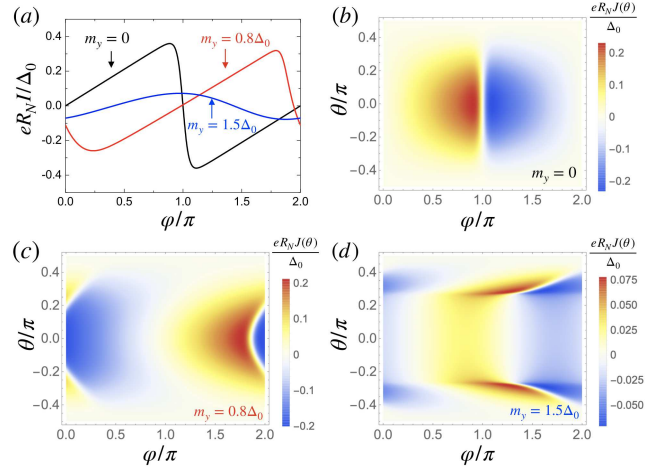}
	\end{center}
	\caption{(a) Current-phase relation for a long SNS junction with $L=5\xi_0$ and $V_G=0$. (b-d) Angle-resolved Josephson current as a function of the incident angle $\theta$ of quasiparticles for $m_y=0$ (b), $m_y=0.8\Delta_0$ (c), and $m_y=1.5\Delta_0$ (d). The temperature is $T=0.01T_c$ and $R_N$ is the resistance when both electrodes are in the normal state. }
	\label{Fig2}
\end{figure}

\emph{Model and Formalism.---}
The superconducting regions are located at $x<0$ and $x>L$, where $L$ is the junction length [Fig. \ref{Fig1}(a)]. An in-plane external magnetic field is applied to the whole structure. The low-energy electronic excitations are described by the Bogoliubov-de Gennes (BdG) Hamiltonian $H=\mathrm{\frac{1}{2}}\sum\nolimits_{\bm{k}}\hat{c}_{\bm{k}}^{\dag}\mathcal{H}_{\bm{k}}\hat{c}_{\bm{k}}$, with
\begin{equation}\label{eq:Hamil}
	\mathcal{H}_{\bm{k}}=vk_{y}\hat{s}_{x}-vk_{x}\hat{s}_{y}\hat{\tau}_{z}-(\mu+V)
	\hat{\tau}_{z}+m_{y}\hat{s}_{y}-\Delta \hat{s}_{y}\hat{\tau}_{y} ,
\end{equation}
and $\hat{c}_{\bm{k}}$ given by $(c_{\bm{k}\uparrow },c_{%
	\bm{k}\downarrow },c_{-\bm{k}\uparrow }^{\dag },c_{-\bm{k}\downarrow }^{\dag
})^{T}$. The chemical potential is $\mu$, the Fermi velocity is $v$, and $\hat{s}_{i}\left( \hat{\tau}%
_{i}\right) $ are Pauli matrices in spin (Nambu) space. The in-plane Zeeman field $m_{y}$ can be induced by either magnetic field or doping of magnetic impurities~\cite{Pablo15}. 
The gate voltage is given by $V(x)=V_G\Theta(|L_N|/2-x)$. 
We consider an induced $s$-wave order parameter $\Delta \left(
x\right) =\Delta \lbrack e^{i\varphi }\Theta \left( -x\right) +\Theta \left(
x-L\right) ]$, where $\varphi $ is the macroscopic superconducting phase difference. We further adopt the BCS temperature dependence $\Delta \left( T\right) =\Delta_{0}\tanh (1.74\sqrt{T_{c}/T-1)}$, with $\Delta _{0}=1.76k_{B}T_{c}$ and $T_{c}$ being the critical temperature. 

The equilibrium Josephson current can be conveniently calculated by Green's functions~\cite{McMillan68}. We construct the retarded Green's function from wavefunctions on the TI, considering the four types of quasiparticle injection processes: left electron (hole) injection,
$\psi_{1}\left( \psi _{2}\right)$, and right electron (hole) injection, $\psi_{3}\left( \psi _{4}\right) $~\cite{McMillan68,Kashiwaya_2000,Pablo15,Lu_2015,Yada15,Lu18}.
Due to translation invariance along the $y$-direction the wavefunctions read $\psi _{i=1- 4}=\psi _{i}\left( x\right) e^{ik_{y}y}$, where $k_{y}=\mu \sin \theta /v$,  with an injection angle $\theta$ for electrons and holes. For $\Delta_0 \ll \mu$ and weak Zeeman coupling $\left\vert m_{y}\right\vert \ll \mu $, $\psi_{1(2)}\left( x\right) $ are taken as
\begin{subequations}
	\begin{alignat}{4}
		\psi _{1}\left( x\right)
		={}&A_{1}e^{ik_{x}x}+a_{1}A_{4}e^{ik_{x}x}+b_{1}A_{3}e^{-ik_{x}x}, 
		\\
		\psi _{2}\left( x\right)
		={}&A_{2}e^{-ik_{x}x}+a_{2}A_{3}e^{-ik_{x}x}+b_{2}A_{4}e^{ik_{x}x}, 
	\end{alignat}%
\end{subequations}
where $k_{x}=\mu \cos \theta /v$. $A_{i}$ are 4-component spinors that can be found in the Supplementary Material (SM)~\cite{SM}. 
The retarded Green's function is obtained by combing all the injection processes~\cite{SM} and defines the extended Furusaki-Tsukada (FT) formula~\cite{FT91} $I(\varphi)=\int d\theta J\left( \theta, \varphi \right) $, with $J\left(  \theta, \varphi \right) =J_{m}\left(  \theta, \varphi \right) +J_{a}\left( \theta, \varphi \right) $ and 
\begin{subequations}
	\begin{alignat}{4}
		J_{m}\left(  \theta, \varphi \right) & =\frac{ek_{B}T}{2\pi }\sum\limits_{\omega _{n}}%
		\left[ \frac{i\Gamma _{+}^{2}}{1-\Gamma _{+}^{2}}-\frac{i\Gamma _{-}^{2}}{%
			1-\Gamma _{-}^{2}}\right] \cos \theta , & & & & & & \\
		J_{a}\left(  \theta, \varphi \right) & =\frac{ek_{B}T}{2\pi }\sum\limits_{\omega _{n}}%
		\left[ \frac{i\Gamma _{+}a_{1}}{1-\Gamma _{+}^{2}}-\frac{i\Gamma _{-}a_{2}}{%
			1-\Gamma _{-}^{2}}\right] \cos \theta . & & & & & &
	\end{alignat}
\end{subequations}
Here, $\Gamma _{\pm }\left( E\right) =\Delta_{0} \left[ E_{\pm }+\sqrt{E_{\pm }^{2}-\Delta _{0}^{2}} \right] ^{-1}$, with $E_{\pm }=E\pm m_{y}\cos \theta $ shows the angle-dependent modification of $a_{1\left( 2\right) }(\omega _{n})$ and $\Gamma _{\pm }\left( \omega_{n}\right) $, which are obtained by analytical continuation, $E\rightarrow i\omega _{n}$, into Matsubara frequencies $\omega _{n}=\pi k_{B}T(2n+1)$, $n=0,\pm 1,\pm 2...$
$J_{m}$ comes from the Doppler shift~\cite{Tanaka02,Tkachov15,Pablo15} of the Fermi surface by $m_{y}$, and $J_{a}$ stems from Andreev reflections. It is straightforward to check that the obtained current formula is consistent with previous results on TI surface in the absence of the in-plane magnetic field~\cite{Tanaka09}. 
	
To compare with the short 1D Josephson junction, we solve the Andreev bound states (ABSs) for arbitrary $L$ and $\theta$ as follows~\cite{SM},
\begin{subequations}
		\begin{alignat}{4}
			\frac{2E_{-}L}{v\cos \theta }+2\arcsin \frac{E_{-}}{\Delta _{0}}& =\left(
			2n+1\right) \pi +\varphi ,  \\
			\frac{2E_{+}L}{v\cos \theta }+2\arcsin \frac{E_{+}}{\Delta _{0}}& =-\varphi
			-\left( 2m+1\right) \pi , 
\end{alignat}
\end{subequations}
with $n$ and $m$ integers. 
While the short 1D junction features one pair of ABSs with constant energy shift $m_y$ ($\cos\theta\rightarrow 1$), in the long junction there are several ABSs present in the spectrum. 
The value of $\varphi $ where the crossing of ABSs occurs is modified by the magnetization, which becomes a function of the angle $\theta$. 
As a result, the current-phase relation (CPR) also becomes angle-dependent, as shown in \cref{Fig2} and explained in detail below. 

\emph{Current-phase relation.---} Using the FT formula, we calculate the CPR of the long junction with a fixed length $L=5\xi _{0}$ and at temperature $T=0.01T_{c}$. 
For our calculations, we choose relevant accessible parameters: The Fermi energy for 3D TIs based on Bi compounds~\cite{Xu2014} can be larger than $50$ meV. The proximitized superconducting gap at zero magnetic field is around $0.5$ meV, and similar values are reported for induced Zeeman exchange from in-plane magnetic fields~\cite{JinFeng20}. 

In~\cref{Fig2}(a), for $m_{y}=0$ the CPR is almost sinusoidal, and as $m_{y}$ increases, the characteristic behavior of CPR becomes anomalous with a finite Josephson current appears for $\varphi =0$. Meanwhile, the maximum current in the CPR, i.e., the critical current $I_{c}$, decreases as $m_{y}$ increases. 
To elucidate the influence of the magnetization on the CPR, we plot the angle-resolved Josephson current in \cref{Fig2}(b-d) corresponding to the
three curves in \cref{Fig2}(a). Clearly, changing $\theta $ shifts the $\varphi $-position of the sign changes in $J( \theta, \varphi)$. 
This is due to the Doppler shift in the coherence factor $\Gamma _{\pm }$, i.e., $E\rightarrow $ $E\pm m_{y}\cos \theta $, as explained above~\cite{SM}. 
Our result coincides with the 1D quantum spin Hall system if we fix $\theta =0$~\cite{Meyer15}. While the Josephson current has non-zero quality factor for a fixed angle $\theta$, the maximum (minimum) current $J( \theta, \varphi )$ appears at different $\varphi $ as $\theta $ varies, see Figs. \ref{Fig2}(b-d). Especially, $J( \theta, \varphi )$ experiences a positive-negative transition which greatly modifies the resultant $I_{c}^{\pm }$ after integration over $\theta $. 
Since the value of $\varphi $ for which the current is maximum (minimum) varies with $\theta$ [\cref{Fig2}(b-d)], we observe a different JDE between the 2D and 1D cases~\footnote{
	\clr{Note that previous works, like, e.g., Refs.~\onlinecite{He_2022,Yuan22}, propose that the diode effect is the result of Cooper pairs acquiring a finite momentum that, in principle, would be the same for all transport modes. In such a case, the sign of the quality factor does not change, even for long 2D junctions. The Doppler shift, by contrast, becomes angle-dependent thus allowing the angle-averaged quality factor to change sign}
}. 

\begin{figure}[t]
	\begin{center}
		\includegraphics[width=85mm]{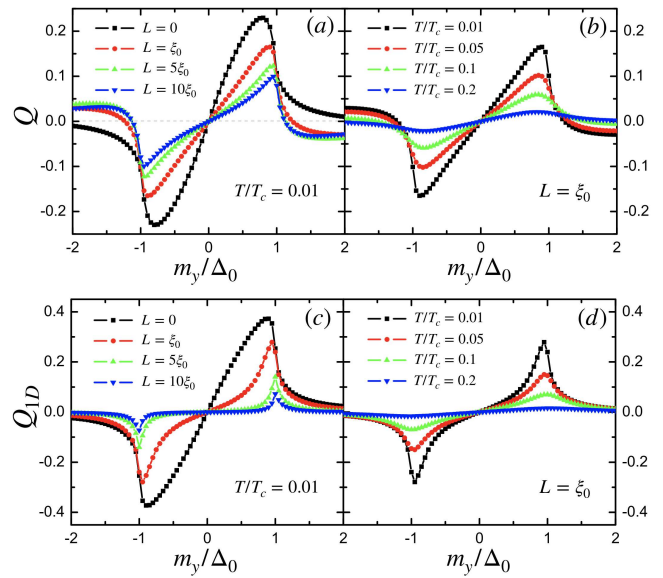}
	\end{center}
	\caption{Quality factor of a 2D SNS junction as a function of the magnetization $m_y$ for several (a) lengths $L$, with fixed temperature $T=0.01T_c$, and (b) temperatures for a fixed length $L=\xi_0$. (c,d) Same for the 1D SNS junction. In all cases, $V_G=0$.} 
	\label{Fig3}
\end{figure}

\emph{Quality factor.---}  We plot in \cref{Fig3}(a) the diode quality factor $Q$ as a function of $m_{y}$ for different lengths of the 2D SNS junction at low temperature, $T=0.01T_{c}$. 
The symmetry relation $Q\left(m_{y}\right) =-$ $Q\left( -m_{y}\right) $ is satisfied and $Q$ depends strongly on the magnitude of the magnetization $m_{y}$. 
The diode effect first grows linearly with increasing $\left\vert m_{y}\right\vert $ and then decreases as $m_{y}$ is larger than some critical value within the gap. Similar behavior is found in 1D SNS junctions~\cite{Davydova22}.
However, we find that the quality factor $Q$ has an additional sign change as $\left\vert m_{y}\right\vert >\Delta _{0}$. In the short junction limit, i.e., $L=0$, such additional sign change disappears. 
In \cref{Fig3}(b), we show the quality factor $Q$ for a long junction at different temperatures ($L=\xi _{0}$). The diode effect is more prominent in the low temperature limit. 

For comparison, we plot the 1D SNS junction by considering the current for $\theta =0$ in \cref{Fig3}(c) and (d). We recover the previous results in Refs.~\onlinecite{Meyer15,Davydova22}, where the quality factor $Q_{1D}$ has no additional sign change as $\left\vert m_{y}\right\vert >\Delta _{0}$, for both short and long junctions. 
We note that the additional sign change has been observed in recent experiments for Dirac systems~\cite{Pal22}. We view this phenomenon as a result of the sign change of the angle-resolved Josephson current for long junctions, cf. \cref{Fig2}. 
Moreover, averaging the Josephson current over $\theta $ combines modes with different signs of the current $J\left( \theta, \varphi \right) $. Such suppression of the resulting Josephson current decreases the value of the diode effect, $\left\vert Q\right\vert <\left\vert Q_{1D}\right\vert $. 

\begin{figure}[tb]
	\begin{center}
		\includegraphics[width=88mm]{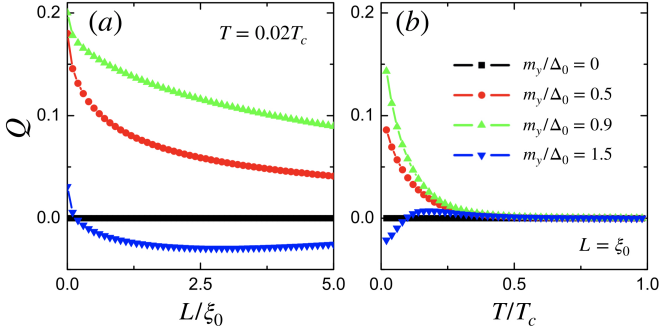}
	\end{center}
	\caption{(a) Quality factor $Q$ as a function of (a) the junction length at $T=0.02T_c$ and (b) the temperature for $L=\xi_0$. The magnetization is $m_y/\Delta_0=0$ (black), $0.5$ (red), $0.9$ (green), and $1.5$ (blue). $V_G=0$ in all panels. }
	\label{Fig4}
\end{figure}

To further illustrate the evolution of the quality factor with the variation of parameters, we show the dependence of $Q$ as a function of junction length and temperature in \cref{Fig4}. One can see that the quality factor for $m_{y}>\Delta_{0}$ (e.g., $m_{y}=1.5\Delta _{0}$) has non-monotonic behavior which can change sign with either $L$ or $T$. For high temperature, the sign of $Q$ is the same for different $m_{y}$ but the magnitude of $Q$ is greatly suppressed. 

\emph{Symmetry analysis.---} We now discuss a symmetry constraint on the
emergence of Josephson diode effect based on our model. From the BdG equations, $\mathcal{H}(\varphi )\psi _{\nu }=E_{\nu }(\varphi )\psi _{\nu }$, 
with eigenstates $E_{\nu}(\varphi )$, the Josephson current reads
\begin{equation}
J(\varphi )=\frac{2e}{h}\partial _{\varphi }F(\varphi ),\quad F(\varphi
)=\sum_{\nu }E_{\nu }(\varphi )f_{\nu },
\end{equation}%
where $f_{\nu }=(1/2)[1-\mathrm{tanh}(E_{\nu }/2k_{B}T)]$ is the Fermi
distribution function. We now assume that the Hamiltonian $\mathcal{H}%
(\varphi )$ preserves a symmetry defined by $\mathcal{U}\mathcal{H}(\varphi )%
\mathcal{U}^{\dagger }=\mathcal{H}(-\varphi )$, which leads to $\mathcal{H}%
(-\varphi )\mathcal{U}\psi _{\nu }=E_{\nu }(\varphi )\mathcal{U}\psi _{\nu }$%
. Thus, both $\mathcal{H}(\varphi )$ and $\mathcal{H}(-\varphi )$ have energy eigenvalues $E_{\nu }(\varphi )$. As a result, in the presence of the symmetry of $\mathcal{U}$, we obtain that the free energy $F$ is an even-function of $\varphi $ and therefore $J(\varphi )$ is an odd-function of $\varphi $%
\begin{equation}
F(\varphi )=F(-\varphi ),\quad J(\varphi )=-J(-\varphi ).
\end{equation}
Since the Josephson current becomes an odd function with respect to $\varphi$, the JDE never occurs in the presence of the symmetry of $\mathcal{U}$. 

We now focus on the symmetry properties of the specific Josephson junction fabricated on a TI. In the absence of Zeeman fields, the Hamiltonian preserves the following symmetries: time-reversal symmetry, $\check{T}\check{H}_{D}{(\boldsymbol{r},\varphi )}\check{T}^{\dagger }=\check{H}_{D}{(\boldsymbol{r},-\varphi )}$, and magnetic-mirror-reflection symmetry $\check{T}_{M}$ with respect to the $zx$-plane, $\check{T%
}_{M}\check{H}_{D}{(\boldsymbol{r},\varphi )}\check{T}_{M}^{\dagger }=\check{%
H}_{D}{(\boldsymbol{r},-\varphi )}$, where $\check{T}$ and $\check{T}_M = \check{M}_{zx}\check{T}$
are
\begin{equation}
\check{T}=\left[
\begin{array}{cc}
	\hat{s}_{y} & 0 \\
	0 & \hat{s}_{y}%
\end{array}%
\right] \mathcal{K}, \quad \check{M}_{zx}=\left[
\begin{array}{cc}
	i\hat{s}_{y} & 0 \\
	0 & i\hat{s}_{y}%
\end{array}%
\right] \mathcal{R}_{y}.
\end{equation}%
Here, $\mathcal{K}$ is the complex conjugation operator and $\mathcal{R}_{y}$ describes the spatial reflection with respect to the $zx$-plane, i.e., $\mathcal{R}_{y}$ converts $y$ into $-y$. As discussed above, these symmetries prohibit the emergence of the JDE. Nevertheless, when we apply a Zeeman field along the $y$-direction, which is described by $\mathcal{H}_{Z,y}=m_{y}\hat{s}_{y}$, both symmetries are broken. As a result, we can expect the Josephson diode effect with the Zeeman field along the $y$-direction. On the other hand, when we apply a Zeeman field in the $x$-direction, the Hamiltonian $\mathcal{H}_{Z,x}=m_{x}\hat{s}_{x}\hat{\tau}_{z}$ still preserves the $\check{T}_{M}$ symmetry. Therefore, we can not obtain the Josephson diode effect with applying the Zeeman field in the $x$-direction. 

\begin{figure}[t]
	\begin{center}
		\includegraphics[width=86mm]{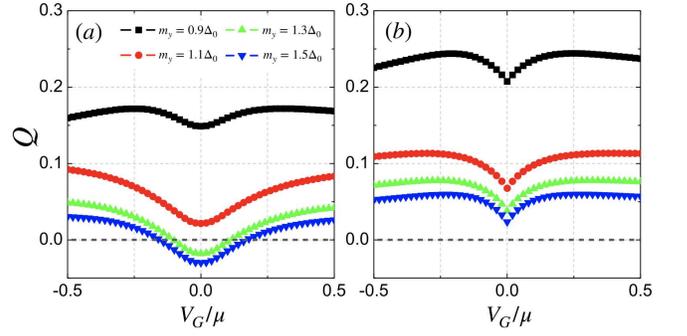}
	\end{center}
	\caption{Quality factor $Q$ as a function of the voltage gate $V_G$ for the (a) long junction ($L/\protect\xi _{0}=2$) and (b) short junction ($L/\protect\xi_0=0.01$). In all cases the barrier length is $L_G/\protect \xi _{0}=0.01$. }
	\label{Fig5}
\end{figure}

\emph{Gate-tuned JDE.---}  On the surface of TIs the JDE is not only affected by the different current-phase relation among multiple transport channels induced by the magnetic field [\cref{Fig2}(c,d)], but also by the magnitude of their contribution to the total current. The latter can be tuned by applying a gate-voltage in the normal region~\cite{Tkachov13}, see \cref{Fig1}(a). 
We thus add a potential barrier with a fixed width $L=10^{-2}\xi _{0}$ inside the normal region where the barrier height $V_{G}$ can be tuned experimentally by changing a gate voltage, see Eq. (\ref{eq:Hamil}). 
For long junctions, \cref{Fig5}(a) shows the dependence of the quality factor on the gate for several magnetic fields. The quality factor is minimal at $V_{G}=0$ and increases with $|V_{G}|$. Interestingly, for large enough magnetic fields, e.g., $m_{y}=1.5\Delta _{0}$, there are sign changes of $Q$ by changing $|V_{G}|$. By contrast, such sign changes are not present in the short junction, see Fig. \ref{Fig5}(b). 
This can be intuitively understood from the fact that a barrier alters the transmissivity of all transport channels by a different amount. Modes with normal incidence are immune to such barrier due to the Klein tunneling, however, the oblique transport has lower transmissivity as compared to the transparent case~\cite{Tkachov13}. As $|V_{G}|$ increases, the characteristic of the JDE behaves similarly to the 1D case, since we have seen that the quality factor has different sign for 2D and 1D junctions under large magnetic field $m_{y}$ (Fig. \ref{Fig3}). Our results with potential barrier open a way towards the electric control by gating of both the sign and the magnitude of the quality factor.

\emph{Conclusions.---} We have studied the diode effect in 2D SNS Josephson junctions on the topological insulator surface. We have demonstrated a high tunability of the quality factor $Q$, including its sign, by the external magnetic field, a gate voltage, and the length of the junction. 
This system can thus be exploited to create an all-in-one device for future superconducting rectification. 

Recently, sign reversals of the supercurrent diode effect have been revealed in bulk Rashba-Zeeman superconductors as a function of temperature and magnetic field~\cite{Yanase22}. Such behavior appears due to the momentum-dependence of the condensation energy in the crossover region of helical superconductivity, where the Cooper-pair-momentum drastically changes. 
%
In the long Josephson junction studied here, the spatial phase gradient becomes smooth as compared to the short junction, and we have found that the Josephson diode shares similar sign reversals with the supercurrent one, but resulting from a different current-phase relation of the transverse transport channels. 
It is also important to emphasize that the sign change here is universal in the long junction limit and has been observed in experiments~\cite{Pal22}. One can also use magnetic doping to realize the Zeeman effect for possible future applications without external magnetic fields. The sign reversal of the AC and DC supercurrent diode effect has been recently reported in Rashba Josephson junctions due to a similar multiple-channel mechanism~\cite{Costa22}.

{\itshape Acknowledgment.} B. L. is supported by the National Natural Science Foundation of China (project 11904257) and the Natural Science Foundation of Tianjin (project 20JCQNJC01310). S. I. is supported by the Grant-in-Aid for JSPS Fellows (JSPS KAKENHI Grant No. JP21J00041) and the JSPS Core-to-Core Program (No. JPJSCCA20170002). P. B. is supported  by the Spanish CM ``Talento Program'' project No.~2019-T1/IND-14088 and the Agencia Estatal de Investigaci\'on project No.~PID2020-117992GA-I00 and No.~CNS2022-135950. Y. T. is supported by Scientific Research (A) (KAKENHI Grant No. JP20H00131), Scientific Research (B) (KAKENHI Grants No. JP20H01857) and JSPS Core-to-Core program Oxide Superspin international network (Grants No. JPJSCCA20170002). N. N. is supported by JST CREST Grant Number JPMJCR1874, Japan, and JSPS KAKENHI Grant number 18H03676.


\bibliography{josephson}

\begin{thebibliography}{78}%
\makeatletter
\providecommand \@ifxundefined [1]{%
 \@ifx{#1\undefined}
}%
\providecommand \@ifnum [1]{%
 \ifnum #1\expandafter \@firstoftwo
 \else \expandafter \@secondoftwo
 \fi
}%
\providecommand \@ifx [1]{%
 \ifx #1\expandafter \@firstoftwo
 \else \expandafter \@secondoftwo
 \fi
}%
\providecommand \natexlab [1]{#1}%
\providecommand \enquote  [1]{``#1''}%
\providecommand \bibnamefont  [1]{#1}%
\providecommand \bibfnamefont [1]{#1}%
\providecommand \citenamefont [1]{#1}%
\providecommand \href@noop [0]{\@secondoftwo}%
\providecommand \href [0]{\begingroup \@sanitize@url \@href}%
\providecommand \@href[1]{\@@startlink{#1}\@@href}%
\providecommand \@@href[1]{\endgroup#1\@@endlink}%
\providecommand \@sanitize@url [0]{\catcode `\\12\catcode `\$12\catcode
  `\&12\catcode `\#12\catcode `\^12\catcode `\_12\catcode `\%12\relax}%
\providecommand \@@startlink[1]{}%
\providecommand \@@endlink[0]{}%
\providecommand \url  [0]{\begingroup\@sanitize@url \@url }%
\providecommand \@url [1]{\endgroup\@href {#1}{\urlprefix }}%
\providecommand \urlprefix  [0]{URL }%
\providecommand \Eprint [0]{\href }%
\providecommand \doibase [0]{http://dx.doi.org/}%
\providecommand \selectlanguage [0]{\@gobble}%
\providecommand \bibinfo  [0]{\@secondoftwo}%
\providecommand \bibfield  [0]{\@secondoftwo}%
\providecommand \translation [1]{[#1]}%
\providecommand \BibitemOpen [0]{}%
\providecommand \bibitemStop [0]{}%
\providecommand \bibitemNoStop [0]{.\EOS\space}%
\providecommand \EOS [0]{\spacefactor3000\relax}%
\providecommand \BibitemShut  [1]{\csname bibitem#1\endcsname}%
\let\auto@bib@innerbib\@empty
\bibitem [{\citenamefont {Rikken}\ and\ \citenamefont
  {Raupach}(1997)}]{Rikken1997}%
  \BibitemOpen
  \bibfield  {author} {\bibinfo {author} {\bibfnamefont {G.~L. J.~A.}\
  \bibnamefont {Rikken}}\ and\ \bibinfo {author} {\bibfnamefont
  {E.}~\bibnamefont {Raupach}},\ }\href {\doibase 10.1038/37323} {\bibfield
  {journal} {\bibinfo  {journal} {Nature}\ }\textbf {\bibinfo {volume} {390}},\
  \bibinfo {pages} {493} (\bibinfo {year} {1997})}\BibitemShut {NoStop}%
\bibitem [{\citenamefont {Rikken}\ \emph {et~al.}(2001)\citenamefont {Rikken},
  \citenamefont {F\"olling},\ and\ \citenamefont {Wyder}}]{Rikken01}%
  \BibitemOpen
  \bibfield  {author} {\bibinfo {author} {\bibfnamefont {G.~L. J.~A.}\
  \bibnamefont {Rikken}}, \bibinfo {author} {\bibfnamefont {J.}~\bibnamefont
  {F\"olling}}, \ and\ \bibinfo {author} {\bibfnamefont {P.}~\bibnamefont
  {Wyder}},\ }\href {\doibase 10.1103/PhysRevLett.87.236602} {\bibfield
  {journal} {\bibinfo  {journal} {Phys. Rev. Lett.}\ }\textbf {\bibinfo
  {volume} {87}},\ \bibinfo {pages} {236602} (\bibinfo {year}
  {2001})}\BibitemShut {NoStop}%
\bibitem [{\citenamefont {Krstic}\ \emph {et~al.}(2002)\citenamefont {Krstic},
  \citenamefont {Roth}, \citenamefont {Burghard}, \citenamefont {Kern},\ and\
  \citenamefont {Rikken}}]{Rikken02}%
  \BibitemOpen
  \bibfield  {author} {\bibinfo {author} {\bibfnamefont {V.}~\bibnamefont
  {Krstic}}, \bibinfo {author} {\bibfnamefont {S.}~\bibnamefont {Roth}},
  \bibinfo {author} {\bibfnamefont {M.}~\bibnamefont {Burghard}}, \bibinfo
  {author} {\bibfnamefont {K.}~\bibnamefont {Kern}}, \ and\ \bibinfo {author}
  {\bibfnamefont {G.~L. J.~A.}\ \bibnamefont {Rikken}},\ }\href {\doibase
  10.1063/1.1523895} {\bibfield  {journal} {\bibinfo  {journal} {The Journal of
  Chemical Physics}\ }\textbf {\bibinfo {volume} {117}},\ \bibinfo {pages}
  {11315} (\bibinfo {year} {2002})}\BibitemShut {NoStop}%
\bibitem [{\citenamefont {Rikken}\ and\ \citenamefont
  {Wyder}(2005)}]{Rikken05}%
  \BibitemOpen
  \bibfield  {author} {\bibinfo {author} {\bibfnamefont {G.~L. J.~A.}\
  \bibnamefont {Rikken}}\ and\ \bibinfo {author} {\bibfnamefont
  {P.}~\bibnamefont {Wyder}},\ }\href {\doibase 10.1103/PhysRevLett.94.016601}
  {\bibfield  {journal} {\bibinfo  {journal} {Phys. Rev. Lett.}\ }\textbf
  {\bibinfo {volume} {94}},\ \bibinfo {pages} {016601} (\bibinfo {year}
  {2005})}\BibitemShut {NoStop}%
\bibitem [{\citenamefont {Pop}\ \emph {et~al.}(2014)\citenamefont {Pop},
  \citenamefont {Auban-Senzier}, \citenamefont {Canadell}, \citenamefont
  {Rikken},\ and\ \citenamefont {Avarvari}}]{Pop2014}%
  \BibitemOpen
  \bibfield  {author} {\bibinfo {author} {\bibfnamefont {F.}~\bibnamefont
  {Pop}}, \bibinfo {author} {\bibfnamefont {P.}~\bibnamefont {Auban-Senzier}},
  \bibinfo {author} {\bibfnamefont {E.}~\bibnamefont {Canadell}}, \bibinfo
  {author} {\bibfnamefont {G.~L. J.~A.}\ \bibnamefont {Rikken}}, \ and\
  \bibinfo {author} {\bibfnamefont {N.}~\bibnamefont {Avarvari}},\ }\href
  {\doibase 10.1038/ncomms4757} {\bibfield  {journal} {\bibinfo  {journal}
  {Nature Communications}\ }\textbf {\bibinfo {volume} {5}},\ \bibinfo {pages}
  {3757} (\bibinfo {year} {2014})}\BibitemShut {NoStop}%
\bibitem [{\citenamefont {Morimoto}\ and\ \citenamefont
  {Nagaosa}(2016)}]{Morimoto16}%
  \BibitemOpen
  \bibfield  {author} {\bibinfo {author} {\bibfnamefont {T.}~\bibnamefont
  {Morimoto}}\ and\ \bibinfo {author} {\bibfnamefont {N.}~\bibnamefont
  {Nagaosa}},\ }\href {\doibase 10.1103/PhysRevLett.117.146603} {\bibfield
  {journal} {\bibinfo  {journal} {Phys. Rev. Lett.}\ }\textbf {\bibinfo
  {volume} {117}},\ \bibinfo {pages} {146603} (\bibinfo {year}
  {2016})}\BibitemShut {NoStop}%
\bibitem [{\citenamefont {Ando}\ \emph {et~al.}(2020)\citenamefont {Ando},
  \citenamefont {Miyasaka}, \citenamefont {Li}, \citenamefont {Ishizuka},
  \citenamefont {Arakawa}, \citenamefont {Shiota}, \citenamefont {Moriyama},
  \citenamefont {Yanase},\ and\ \citenamefont {Ono}}]{OnoNat}%
  \BibitemOpen
  \bibfield  {author} {\bibinfo {author} {\bibfnamefont {F.}~\bibnamefont
  {Ando}}, \bibinfo {author} {\bibfnamefont {Y.}~\bibnamefont {Miyasaka}},
  \bibinfo {author} {\bibfnamefont {T.}~\bibnamefont {Li}}, \bibinfo {author}
  {\bibfnamefont {J.}~\bibnamefont {Ishizuka}}, \bibinfo {author}
  {\bibfnamefont {T.}~\bibnamefont {Arakawa}}, \bibinfo {author} {\bibfnamefont
  {Y.}~\bibnamefont {Shiota}}, \bibinfo {author} {\bibfnamefont
  {T.}~\bibnamefont {Moriyama}}, \bibinfo {author} {\bibfnamefont
  {Y.}~\bibnamefont {Yanase}}, \ and\ \bibinfo {author} {\bibfnamefont
  {T.}~\bibnamefont {Ono}},\ }\href {\doibase 10.1038/s41586-020-2590-4}
  {\bibfield  {journal} {\bibinfo  {journal} {Nature}\ }\textbf {\bibinfo
  {volume} {584}},\ \bibinfo {pages} {373} (\bibinfo {year}
  {2020})}\BibitemShut {NoStop}%
\bibitem [{\citenamefont {Miyasaka}\ \emph {et~al.}(2021)\citenamefont
  {Miyasaka}, \citenamefont {Kawarazaki}, \citenamefont {Narita}, \citenamefont
  {Ando}, \citenamefont {Ikeda}, \citenamefont {Hisatomi}, \citenamefont
  {Daido}, \citenamefont {Shiota}, \citenamefont {Moriyama}, \citenamefont
  {Yanase},\ and\ \citenamefont {Ono}}]{Miyasaka_2021}%
  \BibitemOpen
  \bibfield  {author} {\bibinfo {author} {\bibfnamefont {Y.}~\bibnamefont
  {Miyasaka}}, \bibinfo {author} {\bibfnamefont {R.}~\bibnamefont
  {Kawarazaki}}, \bibinfo {author} {\bibfnamefont {H.}~\bibnamefont {Narita}},
  \bibinfo {author} {\bibfnamefont {F.}~\bibnamefont {Ando}}, \bibinfo {author}
  {\bibfnamefont {Y.}~\bibnamefont {Ikeda}}, \bibinfo {author} {\bibfnamefont
  {R.}~\bibnamefont {Hisatomi}}, \bibinfo {author} {\bibfnamefont
  {A.}~\bibnamefont {Daido}}, \bibinfo {author} {\bibfnamefont
  {Y.}~\bibnamefont {Shiota}}, \bibinfo {author} {\bibfnamefont
  {T.}~\bibnamefont {Moriyama}}, \bibinfo {author} {\bibfnamefont
  {Y.}~\bibnamefont {Yanase}}, \ and\ \bibinfo {author} {\bibfnamefont
  {T.}~\bibnamefont {Ono}},\ }\href {\doibase 10.35848/1882-0786/ac03c0}
  {\bibfield  {journal} {\bibinfo  {journal} {Applied Physics Express}\
  }\textbf {\bibinfo {volume} {14}},\ \bibinfo {pages} {073003} (\bibinfo
  {year} {2021})}\BibitemShut {NoStop}%
\bibitem [{\citenamefont {Zhang}\ \emph {et~al.}(2020)\citenamefont {Zhang},
  \citenamefont {Xu}, \citenamefont {Zou}, \citenamefont {Ai}, \citenamefont
  {Dong}, \citenamefont {Huang}, \citenamefont {Leng}, \citenamefont {Liu},
  \citenamefont {Zhang}, \citenamefont {Jia}, \citenamefont {Peng},
  \citenamefont {Zhao}, \citenamefont {Yang}, \citenamefont {Li}, \citenamefont
  {Guo}, \citenamefont {Haigh}, \citenamefont {Nagaosa}, \citenamefont {Shen},\
  and\ \citenamefont {Xiu}}]{zhang2020}%
  \BibitemOpen
  \bibfield  {author} {\bibinfo {author} {\bibfnamefont {E.}~\bibnamefont
  {Zhang}}, \bibinfo {author} {\bibfnamefont {X.}~\bibnamefont {Xu}}, \bibinfo
  {author} {\bibfnamefont {Y.-C.}\ \bibnamefont {Zou}}, \bibinfo {author}
  {\bibfnamefont {L.}~\bibnamefont {Ai}}, \bibinfo {author} {\bibfnamefont
  {X.}~\bibnamefont {Dong}}, \bibinfo {author} {\bibfnamefont {C.}~\bibnamefont
  {Huang}}, \bibinfo {author} {\bibfnamefont {P.}~\bibnamefont {Leng}},
  \bibinfo {author} {\bibfnamefont {S.}~\bibnamefont {Liu}}, \bibinfo {author}
  {\bibfnamefont {Y.}~\bibnamefont {Zhang}}, \bibinfo {author} {\bibfnamefont
  {Z.}~\bibnamefont {Jia}}, \bibinfo {author} {\bibfnamefont {X.}~\bibnamefont
  {Peng}}, \bibinfo {author} {\bibfnamefont {M.}~\bibnamefont {Zhao}}, \bibinfo
  {author} {\bibfnamefont {Y.}~\bibnamefont {Yang}}, \bibinfo {author}
  {\bibfnamefont {Z.}~\bibnamefont {Li}}, \bibinfo {author} {\bibfnamefont
  {H.}~\bibnamefont {Guo}}, \bibinfo {author} {\bibfnamefont {S.~J.}\
  \bibnamefont {Haigh}}, \bibinfo {author} {\bibfnamefont {N.}~\bibnamefont
  {Nagaosa}}, \bibinfo {author} {\bibfnamefont {J.}~\bibnamefont {Shen}}, \
  and\ \bibinfo {author} {\bibfnamefont {F.}~\bibnamefont {Xiu}},\ }\href
  {\doibase 10.1038/s41467-020-19459-5} {\bibfield  {journal} {\bibinfo
  {journal} {Nature Communications}\ }\textbf {\bibinfo {volume} {11}},\
  \bibinfo {pages} {5634} (\bibinfo {year} {2020})}\BibitemShut {NoStop}%
\bibitem [{\citenamefont {Wakatsuki}\ \emph {et~al.}(2017)\citenamefont
  {Wakatsuki}, \citenamefont {Saito}, \citenamefont {Hoshino}, \citenamefont
  {Itahashi}, \citenamefont {Ideue}, \citenamefont {Ezawa}, \citenamefont
  {Iwasa},\ and\ \citenamefont {Nagaosa}}]{Wakatsuki17}%
  \BibitemOpen
  \bibfield  {author} {\bibinfo {author} {\bibfnamefont {R.}~\bibnamefont
  {Wakatsuki}}, \bibinfo {author} {\bibfnamefont {Y.}~\bibnamefont {Saito}},
  \bibinfo {author} {\bibfnamefont {S.}~\bibnamefont {Hoshino}}, \bibinfo
  {author} {\bibfnamefont {Y.}~\bibnamefont {Itahashi}}, \bibinfo {author}
  {\bibfnamefont {T.}~\bibnamefont {Ideue}}, \bibinfo {author} {\bibfnamefont
  {M.}~\bibnamefont {Ezawa}}, \bibinfo {author} {\bibfnamefont
  {Y.}~\bibnamefont {Iwasa}}, \ and\ \bibinfo {author} {\bibfnamefont
  {N.}~\bibnamefont {Nagaosa}},\ }\href {\doibase 10.1126/sciadv.1602390}
  {\bibfield  {journal} {\bibinfo  {journal} {Science Advances}\ }\textbf
  {\bibinfo {volume} {3}},\ \bibinfo {pages} {e1602390} (\bibinfo {year}
  {2017})}\BibitemShut {NoStop}%
\bibitem [{\citenamefont {Hoshino}\ \emph {et~al.}(2018)\citenamefont
  {Hoshino}, \citenamefont {Wakatsuki}, \citenamefont {Hamamoto},\ and\
  \citenamefont {Nagaosa}}]{Hoshino18}%
  \BibitemOpen
  \bibfield  {author} {\bibinfo {author} {\bibfnamefont {S.}~\bibnamefont
  {Hoshino}}, \bibinfo {author} {\bibfnamefont {R.}~\bibnamefont {Wakatsuki}},
  \bibinfo {author} {\bibfnamefont {K.}~\bibnamefont {Hamamoto}}, \ and\
  \bibinfo {author} {\bibfnamefont {N.}~\bibnamefont {Nagaosa}},\ }\href
  {\doibase 10.1103/PhysRevB.98.054510} {\bibfield  {journal} {\bibinfo
  {journal} {Phys. Rev. B}\ }\textbf {\bibinfo {volume} {98}},\ \bibinfo
  {pages} {054510} (\bibinfo {year} {2018})}\BibitemShut {NoStop}%
\bibitem [{\citenamefont {Lyu}\ \emph {et~al.}(2021)\citenamefont {Lyu},
  \citenamefont {Jiang}, \citenamefont {Wang}, \citenamefont {Xiao},
  \citenamefont {Dong}, \citenamefont {Chen}, \citenamefont {Milošević},
  \citenamefont {Wang}, \citenamefont {Divan}, \citenamefont {Pearson},
  \citenamefont {Wu}, \citenamefont {Peeters},\ and\ \citenamefont
  {Kwok}}]{lyu_2021}%
  \BibitemOpen
  \bibfield  {author} {\bibinfo {author} {\bibfnamefont {Y.-Y.}\ \bibnamefont
  {Lyu}}, \bibinfo {author} {\bibfnamefont {J.}~\bibnamefont {Jiang}}, \bibinfo
  {author} {\bibfnamefont {Y.-L.}\ \bibnamefont {Wang}}, \bibinfo {author}
  {\bibfnamefont {Z.-L.}\ \bibnamefont {Xiao}}, \bibinfo {author}
  {\bibfnamefont {S.}~\bibnamefont {Dong}}, \bibinfo {author} {\bibfnamefont
  {Q.-H.}\ \bibnamefont {Chen}}, \bibinfo {author} {\bibfnamefont {M.~V.}\
  \bibnamefont {Milošević}}, \bibinfo {author} {\bibfnamefont
  {H.}~\bibnamefont {Wang}}, \bibinfo {author} {\bibfnamefont {R.}~\bibnamefont
  {Divan}}, \bibinfo {author} {\bibfnamefont {J.~E.}\ \bibnamefont {Pearson}},
  \bibinfo {author} {\bibfnamefont {P.}~\bibnamefont {Wu}}, \bibinfo {author}
  {\bibfnamefont {F.~M.}\ \bibnamefont {Peeters}}, \ and\ \bibinfo {author}
  {\bibfnamefont {W.-K.}\ \bibnamefont {Kwok}},\ }\href {\doibase
  10.1038/s41467-021-23077-0} {\bibfield  {journal} {\bibinfo  {journal}
  {Nature Communications}\ }\textbf {\bibinfo {volume} {12}},\ \bibinfo {pages}
  {2703} (\bibinfo {year} {2021})}\BibitemShut {NoStop}%
\bibitem [{\citenamefont {Bauriedl}\ \emph {et~al.}(2022)\citenamefont
  {Bauriedl}, \citenamefont {Bäuml}, \citenamefont {Fuchs}, \citenamefont
  {Baumgartner}, \citenamefont {Paulik}, \citenamefont {Bauer}, \citenamefont
  {Lin}, \citenamefont {Lupton}, \citenamefont {Taniguchi}, \citenamefont
  {Watanabe}, \citenamefont {Strunk},\ and\ \citenamefont
  {Paradiso}}]{bauriedl_2022}%
  \BibitemOpen
  \bibfield  {author} {\bibinfo {author} {\bibfnamefont {L.}~\bibnamefont
  {Bauriedl}}, \bibinfo {author} {\bibfnamefont {C.}~\bibnamefont {Bäuml}},
  \bibinfo {author} {\bibfnamefont {L.}~\bibnamefont {Fuchs}}, \bibinfo
  {author} {\bibfnamefont {C.}~\bibnamefont {Baumgartner}}, \bibinfo {author}
  {\bibfnamefont {N.}~\bibnamefont {Paulik}}, \bibinfo {author} {\bibfnamefont
  {J.~M.}\ \bibnamefont {Bauer}}, \bibinfo {author} {\bibfnamefont {K.-Q.}\
  \bibnamefont {Lin}}, \bibinfo {author} {\bibfnamefont {J.~M.}\ \bibnamefont
  {Lupton}}, \bibinfo {author} {\bibfnamefont {T.}~\bibnamefont {Taniguchi}},
  \bibinfo {author} {\bibfnamefont {K.}~\bibnamefont {Watanabe}}, \bibinfo
  {author} {\bibfnamefont {C.}~\bibnamefont {Strunk}}, \ and\ \bibinfo {author}
  {\bibfnamefont {N.}~\bibnamefont {Paradiso}},\ }\href {\doibase
  10.1038/s41467-022-31954-5} {\bibfield  {journal} {\bibinfo  {journal}
  {Nature Communications}\ }\textbf {\bibinfo {volume} {13}},\ \bibinfo {pages}
  {4266} (\bibinfo {year} {2022})}\BibitemShut {NoStop}%
\bibitem [{\citenamefont {Narita}\ \emph {et~al.}(2022)\citenamefont {Narita},
  \citenamefont {Ishizuka}, \citenamefont {Kawarazaki}, \citenamefont {Kan},
  \citenamefont {Shiota}, \citenamefont {Moriyama}, \citenamefont {Shimakawa},
  \citenamefont {Ognev}, \citenamefont {Samardak}, \citenamefont {Yanase},\
  and\ \citenamefont {Ono}}]{narita_2022}%
  \BibitemOpen
  \bibfield  {author} {\bibinfo {author} {\bibfnamefont {H.}~\bibnamefont
  {Narita}}, \bibinfo {author} {\bibfnamefont {J.}~\bibnamefont {Ishizuka}},
  \bibinfo {author} {\bibfnamefont {R.}~\bibnamefont {Kawarazaki}}, \bibinfo
  {author} {\bibfnamefont {D.}~\bibnamefont {Kan}}, \bibinfo {author}
  {\bibfnamefont {Y.}~\bibnamefont {Shiota}}, \bibinfo {author} {\bibfnamefont
  {T.}~\bibnamefont {Moriyama}}, \bibinfo {author} {\bibfnamefont
  {Y.}~\bibnamefont {Shimakawa}}, \bibinfo {author} {\bibfnamefont {A.~V.}\
  \bibnamefont {Ognev}}, \bibinfo {author} {\bibfnamefont {A.~S.}\ \bibnamefont
  {Samardak}}, \bibinfo {author} {\bibfnamefont {Y.}~\bibnamefont {Yanase}}, \
  and\ \bibinfo {author} {\bibfnamefont {T.}~\bibnamefont {Ono}},\ }\href
  {\doibase 10.1038/s41565-022-01159-4} {\bibfield  {journal} {\bibinfo
  {journal} {Nature Nanotechnology}\ }\textbf {\bibinfo {volume} {17}},\
  \bibinfo {pages} {823} (\bibinfo {year} {2022})}\BibitemShut {NoStop}%
\bibitem [{\citenamefont {Strambini}\ \emph {et~al.}(2022)\citenamefont
  {Strambini}, \citenamefont {Spies}, \citenamefont {Ligato}, \citenamefont
  {Ilić}, \citenamefont {Rouco}, \citenamefont {González-Orellana},
  \citenamefont {Ilyn}, \citenamefont {Rogero}, \citenamefont {Bergeret},
  \citenamefont {Moodera}, \citenamefont {Virtanen}, \citenamefont
  {Heikkilä},\ and\ \citenamefont {Giazotto}}]{strambini_2022}%
  \BibitemOpen
  \bibfield  {author} {\bibinfo {author} {\bibfnamefont {E.}~\bibnamefont
  {Strambini}}, \bibinfo {author} {\bibfnamefont {M.}~\bibnamefont {Spies}},
  \bibinfo {author} {\bibfnamefont {N.}~\bibnamefont {Ligato}}, \bibinfo
  {author} {\bibfnamefont {S.}~\bibnamefont {Ilić}}, \bibinfo {author}
  {\bibfnamefont {M.}~\bibnamefont {Rouco}}, \bibinfo {author} {\bibfnamefont
  {C.}~\bibnamefont {González-Orellana}}, \bibinfo {author} {\bibfnamefont
  {M.}~\bibnamefont {Ilyn}}, \bibinfo {author} {\bibfnamefont {C.}~\bibnamefont
  {Rogero}}, \bibinfo {author} {\bibfnamefont {F.~S.}\ \bibnamefont
  {Bergeret}}, \bibinfo {author} {\bibfnamefont {J.~S.}\ \bibnamefont
  {Moodera}}, \bibinfo {author} {\bibfnamefont {P.}~\bibnamefont {Virtanen}},
  \bibinfo {author} {\bibfnamefont {T.~T.}\ \bibnamefont {Heikkilä}}, \ and\
  \bibinfo {author} {\bibfnamefont {F.}~\bibnamefont {Giazotto}},\ }\href
  {\doibase 10.1038/s41467-022-29990-2} {\bibfield  {journal} {\bibinfo
  {journal} {Nature Communications}\ }\textbf {\bibinfo {volume} {13}},\
  \bibinfo {pages} {2431} (\bibinfo {year} {2022})}\BibitemShut {NoStop}%
\bibitem [{\citenamefont {Karabassov}\ \emph
  {et~al.}(2022{\natexlab{a}})\citenamefont {Karabassov}, \citenamefont
  {Bobkova}, \citenamefont {Golubov},\ and\ \citenamefont
  {Vasenko}}]{Golubov22}%
  \BibitemOpen
  \bibfield  {author} {\bibinfo {author} {\bibfnamefont {T.}~\bibnamefont
  {Karabassov}}, \bibinfo {author} {\bibfnamefont {I.~V.}\ \bibnamefont
  {Bobkova}}, \bibinfo {author} {\bibfnamefont {A.~A.}\ \bibnamefont
  {Golubov}}, \ and\ \bibinfo {author} {\bibfnamefont {A.~S.}\ \bibnamefont
  {Vasenko}},\ }\href {\doibase 10.1103/PhysRevB.106.224509} {\bibfield
  {journal} {\bibinfo  {journal} {Phys. Rev. B}\ }\textbf {\bibinfo {volume}
  {106}},\ \bibinfo {pages} {224509} (\bibinfo {year}
  {2022}{\natexlab{a}})}\BibitemShut {NoStop}%
\bibitem [{\citenamefont {Lin}\ \emph {et~al.}(2022)\citenamefont {Lin},
  \citenamefont {Siriviboon}, \citenamefont {Scammell}, \citenamefont {Liu},
  \citenamefont {Rhodes}, \citenamefont {Watanabe}, \citenamefont {Taniguchi},
  \citenamefont {Hone}, \citenamefont {Scheurer},\ and\ \citenamefont
  {Li}}]{lin_2022}%
  \BibitemOpen
  \bibfield  {author} {\bibinfo {author} {\bibfnamefont {J.-X.}\ \bibnamefont
  {Lin}}, \bibinfo {author} {\bibfnamefont {P.}~\bibnamefont {Siriviboon}},
  \bibinfo {author} {\bibfnamefont {H.}~\bibnamefont {Scammell}}, \bibinfo
  {author} {\bibfnamefont {S.}~\bibnamefont {Liu}}, \bibinfo {author}
  {\bibfnamefont {D.}~\bibnamefont {Rhodes}}, \bibinfo {author} {\bibfnamefont
  {K.}~\bibnamefont {Watanabe}}, \bibinfo {author} {\bibfnamefont
  {T.}~\bibnamefont {Taniguchi}}, \bibinfo {author} {\bibfnamefont
  {J.}~\bibnamefont {Hone}}, \bibinfo {author} {\bibfnamefont {M.}~\bibnamefont
  {Scheurer}}, \ and\ \bibinfo {author} {\bibfnamefont {J.}~\bibnamefont
  {Li}},\ }\href {\doibase 10.1038/s41567-022-01700-1} {\bibfield  {journal}
  {\bibinfo  {journal} {Nature Physics}\ }\textbf {\bibinfo {volume} {18}},\
  \bibinfo {pages} {1221} (\bibinfo {year} {2022})}\BibitemShut {NoStop}%
\bibitem [{\citenamefont {Hu}\ \emph {et~al.}(2022)\citenamefont {Hu},
  \citenamefont {Sun}, \citenamefont {Xie},\ and\ \citenamefont {Law}}]{HuXin}%
  \BibitemOpen
  \bibfield  {author} {\bibinfo {author} {\bibfnamefont {J.-X.}\ \bibnamefont
  {Hu}}, \bibinfo {author} {\bibfnamefont {Z.-T.}\ \bibnamefont {Sun}},
  \bibinfo {author} {\bibfnamefont {Y.-M.}\ \bibnamefont {Xie}}, \ and\
  \bibinfo {author} {\bibfnamefont {K.~T.}\ \bibnamefont {Law}},\ }\href
  {\doibase 10.48550/arXiv.2211.14846} {\enquote {\bibinfo {title} {Valley
  polarization induced josephson diode effect in twisted bilayer graphene},}\ }
  (\bibinfo {year} {2022}),\ \Eprint {http://arxiv.org/abs/2211.14846}
  {arXiv:2211.14846} \BibitemShut {NoStop}%
\bibitem [{\citenamefont {Alvarado}\ \emph {et~al.}(2023)\citenamefont
  {Alvarado}, \citenamefont {Burset},\ and\ \citenamefont
  {Levy~Yeyati}}]{Alvarado23}%
  \BibitemOpen
  \bibfield  {author} {\bibinfo {author} {\bibfnamefont {M.}~\bibnamefont
  {Alvarado}}, \bibinfo {author} {\bibfnamefont {P.}~\bibnamefont {Burset}}, \
  and\ \bibinfo {author} {\bibfnamefont {A.}~\bibnamefont {Levy~Yeyati}},\
  }\href {\doibase 10.48550/arXiv.2303.07738} {\enquote {\bibinfo {title}
  {Intrinsic non-magnetic $\phi_0$ {Josephson} junctions in twisted bilayer
  graphene},}\ } (\bibinfo {year} {2023}),\ \Eprint
  {http://arxiv.org/abs/2303.07738} {arXiv:2303.07738} \BibitemShut {NoStop}%
\bibitem [{\citenamefont {Bocquillon}\ \emph {et~al.}(2016)\citenamefont
  {Bocquillon}, \citenamefont {Deacon}, \citenamefont {Wiedenmann},
  \citenamefont {Leubner}, \citenamefont {Klapwijk}, \citenamefont {Brüne},
  \citenamefont {Ishibashi}, \citenamefont {Buhmann},\ and\ \citenamefont
  {Molenkamp}}]{Bocquillon16}%
  \BibitemOpen
  \bibfield  {author} {\bibinfo {author} {\bibfnamefont {E.}~\bibnamefont
  {Bocquillon}}, \bibinfo {author} {\bibfnamefont {R.}~\bibnamefont {Deacon}},
  \bibinfo {author} {\bibfnamefont {J.}~\bibnamefont {Wiedenmann}}, \bibinfo
  {author} {\bibfnamefont {P.}~\bibnamefont {Leubner}}, \bibinfo {author}
  {\bibfnamefont {T.}~\bibnamefont {Klapwijk}}, \bibinfo {author}
  {\bibfnamefont {C.}~\bibnamefont {Brüne}}, \bibinfo {author} {\bibfnamefont
  {K.}~\bibnamefont {Ishibashi}}, \bibinfo {author} {\bibfnamefont
  {H.}~\bibnamefont {Buhmann}}, \ and\ \bibinfo {author} {\bibfnamefont
  {L.}~\bibnamefont {Molenkamp}},\ }\href {\doibase 10.1038/nnano.2016.159}
  {\bibfield  {journal} {\bibinfo  {journal} {Nature Nanotechnology}\ }\textbf
  {\bibinfo {volume} {12}},\ \bibinfo {pages} {137} (\bibinfo {year}
  {2016})}\BibitemShut {NoStop}%
\bibitem [{\citenamefont {Baumgartner}\ \emph
  {et~al.}(2022{\natexlab{a}})\citenamefont {Baumgartner}, \citenamefont
  {Fuchs}, \citenamefont {Costa}, \citenamefont {Reinhardt}, \citenamefont
  {Gronin}, \citenamefont {Gardner}, \citenamefont {Lindemann}, \citenamefont
  {Manfra}, \citenamefont {Faria~Junior}, \citenamefont {Kochan}, \citenamefont
  {Fabian}, \citenamefont {Paradiso},\ and\ \citenamefont
  {Strunk}}]{Baumgartner22}%
  \BibitemOpen
  \bibfield  {author} {\bibinfo {author} {\bibfnamefont {C.}~\bibnamefont
  {Baumgartner}}, \bibinfo {author} {\bibfnamefont {L.}~\bibnamefont {Fuchs}},
  \bibinfo {author} {\bibfnamefont {A.}~\bibnamefont {Costa}}, \bibinfo
  {author} {\bibfnamefont {S.}~\bibnamefont {Reinhardt}}, \bibinfo {author}
  {\bibfnamefont {S.}~\bibnamefont {Gronin}}, \bibinfo {author} {\bibfnamefont
  {G.}~\bibnamefont {Gardner}}, \bibinfo {author} {\bibfnamefont
  {T.}~\bibnamefont {Lindemann}}, \bibinfo {author} {\bibfnamefont
  {M.}~\bibnamefont {Manfra}}, \bibinfo {author} {\bibfnamefont
  {P.}~\bibnamefont {Faria~Junior}}, \bibinfo {author} {\bibfnamefont
  {D.}~\bibnamefont {Kochan}}, \bibinfo {author} {\bibfnamefont
  {J.}~\bibnamefont {Fabian}}, \bibinfo {author} {\bibfnamefont
  {N.}~\bibnamefont {Paradiso}}, \ and\ \bibinfo {author} {\bibfnamefont
  {C.}~\bibnamefont {Strunk}},\ }\href {\doibase 10.1038/s41565-021-01009-9}
  {\bibfield  {journal} {\bibinfo  {journal} {Nature Nanotechnology}\ }\textbf
  {\bibinfo {volume} {17}},\ \bibinfo {pages} {39} (\bibinfo {year}
  {2022}{\natexlab{a}})}\BibitemShut {NoStop}%
\bibitem [{\citenamefont {Wu}\ \emph {et~al.}(2022)\citenamefont {Wu},
  \citenamefont {Wang}, \citenamefont {Xu}, \citenamefont {Sivakumar},
  \citenamefont {Pasco}, \citenamefont {Filippozzi}, \citenamefont {Parkin},
  \citenamefont {Zeng}, \citenamefont {McQueen},\ and\ \citenamefont
  {Ali}}]{Wu22}%
  \BibitemOpen
  \bibfield  {author} {\bibinfo {author} {\bibfnamefont {H.}~\bibnamefont
  {Wu}}, \bibinfo {author} {\bibfnamefont {Y.}~\bibnamefont {Wang}}, \bibinfo
  {author} {\bibfnamefont {Y.}~\bibnamefont {Xu}}, \bibinfo {author}
  {\bibfnamefont {P.}~\bibnamefont {Sivakumar}}, \bibinfo {author}
  {\bibfnamefont {C.}~\bibnamefont {Pasco}}, \bibinfo {author} {\bibfnamefont
  {U.}~\bibnamefont {Filippozzi}}, \bibinfo {author} {\bibfnamefont
  {S.}~\bibnamefont {Parkin}}, \bibinfo {author} {\bibfnamefont
  {Y.}~\bibnamefont {Zeng}}, \bibinfo {author} {\bibfnamefont {T.}~\bibnamefont
  {McQueen}}, \ and\ \bibinfo {author} {\bibfnamefont {M.}~\bibnamefont
  {Ali}},\ }\href {\doibase 10.1038/s41586-022-04504-8} {\bibfield  {journal}
  {\bibinfo  {journal} {Nature}\ }\textbf {\bibinfo {volume} {604}},\ \bibinfo
  {pages} {653} (\bibinfo {year} {2022})}\BibitemShut {NoStop}%
\bibitem [{\citenamefont {Diez-Merida}\ \emph {et~al.}(2021)\citenamefont
  {Diez-Merida}, \citenamefont {Diez-Carlon}, \citenamefont {Yang},
  \citenamefont {Xie}, \citenamefont {Gao}, \citenamefont {Watanabe},
  \citenamefont {Taniguchi}, \citenamefont {Lu}, \citenamefont {Law},\ and\
  \citenamefont {Efetov}}]{Merida21}%
  \BibitemOpen
  \bibfield  {author} {\bibinfo {author} {\bibfnamefont {J.}~\bibnamefont
  {Diez-Merida}}, \bibinfo {author} {\bibfnamefont {A.}~\bibnamefont
  {Diez-Carlon}}, \bibinfo {author} {\bibfnamefont {S.}~\bibnamefont {Yang}},
  \bibinfo {author} {\bibfnamefont {Y.}~\bibnamefont {Xie}}, \bibinfo {author}
  {\bibfnamefont {X.}~\bibnamefont {Gao}}, \bibinfo {author} {\bibfnamefont
  {K.}~\bibnamefont {Watanabe}}, \bibinfo {author} {\bibfnamefont
  {T.}~\bibnamefont {Taniguchi}}, \bibinfo {author} {\bibfnamefont
  {X.}~\bibnamefont {Lu}}, \bibinfo {author} {\bibfnamefont {K.}~\bibnamefont
  {Law}}, \ and\ \bibinfo {author} {\bibfnamefont {D.}~\bibnamefont {Efetov}},\
  }\href {\doibase https://doi.org/10.48550/arXiv.2110.01067} {\enquote
  {\bibinfo {title} {Magnetic josephson junctions and superconducting diodes in
  magic angle twisted bilayer graphene},}\ } (\bibinfo {year} {2021}),\ \Eprint
  {http://arxiv.org/abs/2110.01067} {arXiv:2110.01067} \BibitemShut {NoStop}%
\bibitem [{\citenamefont {Pal}\ \emph {et~al.}(2022)\citenamefont {Pal},
  \citenamefont {Chakraborty}, \citenamefont {Sivakumar}, \citenamefont
  {Davydova}, \citenamefont {Gopi}, \citenamefont {Pandeya}, \citenamefont
  {Krieger}, \citenamefont {Zhang}, \citenamefont {Date}, \citenamefont {Ju},
  \citenamefont {Yuan}, \citenamefont {Schröter}, \citenamefont {Fu},\ and\
  \citenamefont {Parkin}}]{Pal22}%
  \BibitemOpen
  \bibfield  {author} {\bibinfo {author} {\bibfnamefont {B.}~\bibnamefont
  {Pal}}, \bibinfo {author} {\bibfnamefont {A.}~\bibnamefont {Chakraborty}},
  \bibinfo {author} {\bibfnamefont {P.~K.}\ \bibnamefont {Sivakumar}}, \bibinfo
  {author} {\bibfnamefont {M.}~\bibnamefont {Davydova}}, \bibinfo {author}
  {\bibfnamefont {A.~K.}\ \bibnamefont {Gopi}}, \bibinfo {author}
  {\bibfnamefont {A.~K.}\ \bibnamefont {Pandeya}}, \bibinfo {author}
  {\bibfnamefont {J.~A.}\ \bibnamefont {Krieger}}, \bibinfo {author}
  {\bibfnamefont {Y.}~\bibnamefont {Zhang}}, \bibinfo {author} {\bibfnamefont
  {M.}~\bibnamefont {Date}}, \bibinfo {author} {\bibfnamefont {S.}~\bibnamefont
  {Ju}}, \bibinfo {author} {\bibfnamefont {N.}~\bibnamefont {Yuan}}, \bibinfo
  {author} {\bibfnamefont {N.~B.~M.}\ \bibnamefont {Schröter}}, \bibinfo
  {author} {\bibfnamefont {L.}~\bibnamefont {Fu}}, \ and\ \bibinfo {author}
  {\bibfnamefont {S.~S.~P.}\ \bibnamefont {Parkin}},\ }\href {\doibase
  10.1038/s41567-022-01699-5} {\bibfield  {journal} {\bibinfo  {journal}
  {Nature Physics}\ }\textbf {\bibinfo {volume} {18}},\ \bibinfo {pages} {1228}
  (\bibinfo {year} {2022})}\BibitemShut {NoStop}%
\bibitem [{\citenamefont {Baumgartner}\ \emph
  {et~al.}(2022{\natexlab{b}})\citenamefont {Baumgartner}, \citenamefont
  {Fuchs}, \citenamefont {Costa}, \citenamefont {Pic{\'{o}}-Cort{\'{e}}s},
  \citenamefont {Reinhardt}, \citenamefont {Gronin}, \citenamefont {Gardner},
  \citenamefont {Lindemann}, \citenamefont {Manfra}, \citenamefont {Junior},
  \citenamefont {Kochan}, \citenamefont {Fabian}, \citenamefont {Paradiso},\
  and\ \citenamefont {Strunk}}]{Baumgartner_2022}%
  \BibitemOpen
  \bibfield  {author} {\bibinfo {author} {\bibfnamefont {C.}~\bibnamefont
  {Baumgartner}}, \bibinfo {author} {\bibfnamefont {L.}~\bibnamefont {Fuchs}},
  \bibinfo {author} {\bibfnamefont {A.}~\bibnamefont {Costa}}, \bibinfo
  {author} {\bibfnamefont {J.}~\bibnamefont {Pic{\'{o}}-Cort{\'{e}}s}},
  \bibinfo {author} {\bibfnamefont {S.}~\bibnamefont {Reinhardt}}, \bibinfo
  {author} {\bibfnamefont {S.}~\bibnamefont {Gronin}}, \bibinfo {author}
  {\bibfnamefont {G.~C.}\ \bibnamefont {Gardner}}, \bibinfo {author}
  {\bibfnamefont {T.}~\bibnamefont {Lindemann}}, \bibinfo {author}
  {\bibfnamefont {M.~J.}\ \bibnamefont {Manfra}}, \bibinfo {author}
  {\bibfnamefont {P.~E.~F.}\ \bibnamefont {Junior}}, \bibinfo {author}
  {\bibfnamefont {D.}~\bibnamefont {Kochan}}, \bibinfo {author} {\bibfnamefont
  {J.}~\bibnamefont {Fabian}}, \bibinfo {author} {\bibfnamefont
  {N.}~\bibnamefont {Paradiso}}, \ and\ \bibinfo {author} {\bibfnamefont
  {C.}~\bibnamefont {Strunk}},\ }\href {\doibase 10.1088/1361-648x/ac4d5e}
  {\bibfield  {journal} {\bibinfo  {journal} {Journal of Physics: Condensed
  Matter}\ }\textbf {\bibinfo {volume} {34}},\ \bibinfo {pages} {154005}
  (\bibinfo {year} {2022}{\natexlab{b}})}\BibitemShut {NoStop}%
\bibitem [{\citenamefont {Gupta}\ \emph {et~al.}(2022)\citenamefont {Gupta},
  \citenamefont {Graziano}, \citenamefont {Pendharkar}, \citenamefont {Dong},
  \citenamefont {Dempsey}, \citenamefont {Palmstrøm},\ and\ \citenamefont
  {Pribiag}}]{Gupta22}%
  \BibitemOpen
  \bibfield  {author} {\bibinfo {author} {\bibfnamefont {M.}~\bibnamefont
  {Gupta}}, \bibinfo {author} {\bibfnamefont {G.}~\bibnamefont {Graziano}},
  \bibinfo {author} {\bibfnamefont {M.}~\bibnamefont {Pendharkar}}, \bibinfo
  {author} {\bibfnamefont {J.}~\bibnamefont {Dong}}, \bibinfo {author}
  {\bibfnamefont {C.}~\bibnamefont {Dempsey}}, \bibinfo {author} {\bibfnamefont
  {C.}~\bibnamefont {Palmstrøm}}, \ and\ \bibinfo {author} {\bibfnamefont
  {V.}~\bibnamefont {Pribiag}},\ }\href@noop {} {\enquote {\bibinfo {title}
  {Superconducting diode effect in a three-terminal josephson device},}\ }
  (\bibinfo {year} {2022}),\ \Eprint {http://arxiv.org/abs/2206.08471}
  {arXiv:2206.08471} \BibitemShut {NoStop}%
\bibitem [{\citenamefont {Turini}\ \emph {et~al.}(2022)\citenamefont {Turini},
  \citenamefont {Salimian}, \citenamefont {Carrega}, \citenamefont {Iorio},
  \citenamefont {Strambini}, \citenamefont {Giazotto}, \citenamefont {Zannier},
  \citenamefont {Sorba},\ and\ \citenamefont {Heun}}]{Turini22}%
  \BibitemOpen
  \bibfield  {author} {\bibinfo {author} {\bibfnamefont {B.}~\bibnamefont
  {Turini}}, \bibinfo {author} {\bibfnamefont {S.}~\bibnamefont {Salimian}},
  \bibinfo {author} {\bibfnamefont {M.}~\bibnamefont {Carrega}}, \bibinfo
  {author} {\bibfnamefont {A.}~\bibnamefont {Iorio}}, \bibinfo {author}
  {\bibfnamefont {E.}~\bibnamefont {Strambini}}, \bibinfo {author}
  {\bibfnamefont {F.}~\bibnamefont {Giazotto}}, \bibinfo {author}
  {\bibfnamefont {V.}~\bibnamefont {Zannier}}, \bibinfo {author} {\bibfnamefont
  {L.}~\bibnamefont {Sorba}}, \ and\ \bibinfo {author} {\bibfnamefont
  {S.}~\bibnamefont {Heun}},\ }\href@noop {} {\enquote {\bibinfo {title}
  {Josephson diode effect in high mobility insb nanoflags},}\ } (\bibinfo
  {year} {2022}),\ \Eprint {http://arxiv.org/abs/2207.08772} {arXiv:2207.08772}
  \BibitemShut {NoStop}%
\bibitem [{\citenamefont {Hu}\ \emph {et~al.}(2007)\citenamefont {Hu},
  \citenamefont {Wu},\ and\ \citenamefont {Dai}}]{Jiangping07}%
  \BibitemOpen
  \bibfield  {author} {\bibinfo {author} {\bibfnamefont {J.}~\bibnamefont
  {Hu}}, \bibinfo {author} {\bibfnamefont {C.}~\bibnamefont {Wu}}, \ and\
  \bibinfo {author} {\bibfnamefont {X.}~\bibnamefont {Dai}},\ }\href {\doibase
  10.1103/PhysRevLett.99.067004} {\bibfield  {journal} {\bibinfo  {journal}
  {Phys. Rev. Lett.}\ }\textbf {\bibinfo {volume} {99}},\ \bibinfo {pages}
  {067004} (\bibinfo {year} {2007})}\BibitemShut {NoStop}%
\bibitem [{\citenamefont {Zhang}\ \emph {et~al.}(2022)\citenamefont {Zhang},
  \citenamefont {Gu}, \citenamefont {Li}, \citenamefont {Hu},\ and\
  \citenamefont {Jiang}}]{Zhang21}%
  \BibitemOpen
  \bibfield  {author} {\bibinfo {author} {\bibfnamefont {Y.}~\bibnamefont
  {Zhang}}, \bibinfo {author} {\bibfnamefont {Y.}~\bibnamefont {Gu}}, \bibinfo
  {author} {\bibfnamefont {P.}~\bibnamefont {Li}}, \bibinfo {author}
  {\bibfnamefont {J.}~\bibnamefont {Hu}}, \ and\ \bibinfo {author}
  {\bibfnamefont {K.}~\bibnamefont {Jiang}},\ }\href {\doibase
  10.1103/PhysRevX.12.041013} {\bibfield  {journal} {\bibinfo  {journal} {Phys.
  Rev. X}\ }\textbf {\bibinfo {volume} {12}},\ \bibinfo {pages} {041013}
  (\bibinfo {year} {2022})}\BibitemShut {NoStop}%
\bibitem [{\citenamefont {Misaki}\ and\ \citenamefont
  {Nagaosa}(2021)}]{Misaki21}%
  \BibitemOpen
  \bibfield  {author} {\bibinfo {author} {\bibfnamefont {K.}~\bibnamefont
  {Misaki}}\ and\ \bibinfo {author} {\bibfnamefont {N.}~\bibnamefont
  {Nagaosa}},\ }\href {\doibase 10.1103/PhysRevB.103.245302} {\bibfield
  {journal} {\bibinfo  {journal} {Phys. Rev. B}\ }\textbf {\bibinfo {volume}
  {103}},\ \bibinfo {pages} {245302} (\bibinfo {year} {2021})}\BibitemShut
  {NoStop}%
\bibitem [{\citenamefont {Kopasov}\ \emph {et~al.}(2021)\citenamefont
  {Kopasov}, \citenamefont {Kutlin},\ and\ \citenamefont
  {Mel'nikov}}]{Kopasov21}%
  \BibitemOpen
  \bibfield  {author} {\bibinfo {author} {\bibfnamefont {A.~A.}\ \bibnamefont
  {Kopasov}}, \bibinfo {author} {\bibfnamefont {A.~G.}\ \bibnamefont {Kutlin}},
  \ and\ \bibinfo {author} {\bibfnamefont {A.~S.}\ \bibnamefont {Mel'nikov}},\
  }\href {\doibase 10.1103/PhysRevB.103.144520} {\bibfield  {journal} {\bibinfo
   {journal} {Phys. Rev. B}\ }\textbf {\bibinfo {volume} {103}},\ \bibinfo
  {pages} {144520} (\bibinfo {year} {2021})}\BibitemShut {NoStop}%
\bibitem [{\citenamefont {He}\ \emph {et~al.}(2022)\citenamefont {He},
  \citenamefont {Tanaka},\ and\ \citenamefont {Nagaosa}}]{He_2022}%
  \BibitemOpen
  \bibfield  {author} {\bibinfo {author} {\bibfnamefont {J.~J.}\ \bibnamefont
  {He}}, \bibinfo {author} {\bibfnamefont {Y.}~\bibnamefont {Tanaka}}, \ and\
  \bibinfo {author} {\bibfnamefont {N.}~\bibnamefont {Nagaosa}},\ }\href
  {\doibase 10.1088/1367-2630/ac6766} {\bibfield  {journal} {\bibinfo
  {journal} {New Journal of Physics}\ }\textbf {\bibinfo {volume} {24}},\
  \bibinfo {pages} {053014} (\bibinfo {year} {2022})}\BibitemShut {NoStop}%
\bibitem [{\citenamefont {Daido}\ \emph {et~al.}(2022)\citenamefont {Daido},
  \citenamefont {Ikeda},\ and\ \citenamefont {Yanase}}]{Yanase22}%
  \BibitemOpen
  \bibfield  {author} {\bibinfo {author} {\bibfnamefont {A.}~\bibnamefont
  {Daido}}, \bibinfo {author} {\bibfnamefont {Y.}~\bibnamefont {Ikeda}}, \ and\
  \bibinfo {author} {\bibfnamefont {Y.}~\bibnamefont {Yanase}},\ }\href
  {\doibase 10.1103/PhysRevLett.128.037001} {\bibfield  {journal} {\bibinfo
  {journal} {Phys. Rev. Lett.}\ }\textbf {\bibinfo {volume} {128}},\ \bibinfo
  {pages} {037001} (\bibinfo {year} {2022})}\BibitemShut {NoStop}%
\bibitem [{\citenamefont {Yuan}\ and\ \citenamefont {Fu}(2022)}]{Yuan22}%
  \BibitemOpen
  \bibfield  {author} {\bibinfo {author} {\bibfnamefont {N.~F.~Q.}\
  \bibnamefont {Yuan}}\ and\ \bibinfo {author} {\bibfnamefont {L.}~\bibnamefont
  {Fu}},\ }\href {\doibase 10.1073/pnas.2119548119} {\bibfield  {journal}
  {\bibinfo  {journal} {Proceedings of the National Academy of Sciences}\
  }\textbf {\bibinfo {volume} {119}},\ \bibinfo {pages} {e2119548119} (\bibinfo
  {year} {2022})}\BibitemShut {NoStop}%
\bibitem [{\citenamefont {Ili\ifmmode~\acute{c}\else \'{c}\fi{}}\ and\
  \citenamefont {Bergeret}(2022)}]{Bergeret22}%
  \BibitemOpen
  \bibfield  {author} {\bibinfo {author} {\bibfnamefont {S.}~\bibnamefont
  {Ili\ifmmode~\acute{c}\else \'{c}\fi{}}}\ and\ \bibinfo {author}
  {\bibfnamefont {F.~S.}\ \bibnamefont {Bergeret}},\ }\href {\doibase
  10.1103/PhysRevLett.128.177001} {\bibfield  {journal} {\bibinfo  {journal}
  {Phys. Rev. Lett.}\ }\textbf {\bibinfo {volume} {128}},\ \bibinfo {pages}
  {177001} (\bibinfo {year} {2022})}\BibitemShut {NoStop}%
\bibitem [{\citenamefont {Halterman}\ \emph {et~al.}(2022)\citenamefont
  {Halterman}, \citenamefont {Alidoust}, \citenamefont {Smith},\ and\
  \citenamefont {Starr}}]{Halterman22}%
  \BibitemOpen
  \bibfield  {author} {\bibinfo {author} {\bibfnamefont {K.}~\bibnamefont
  {Halterman}}, \bibinfo {author} {\bibfnamefont {M.}~\bibnamefont {Alidoust}},
  \bibinfo {author} {\bibfnamefont {R.}~\bibnamefont {Smith}}, \ and\ \bibinfo
  {author} {\bibfnamefont {S.}~\bibnamefont {Starr}},\ }\href {\doibase
  10.1103/PhysRevB.105.104508} {\bibfield  {journal} {\bibinfo  {journal}
  {Phys. Rev. B}\ }\textbf {\bibinfo {volume} {105}},\ \bibinfo {pages}
  {104508} (\bibinfo {year} {2022})}\BibitemShut {NoStop}%
\bibitem [{\citenamefont {Karabassov}\ \emph
  {et~al.}(2022{\natexlab{b}})\citenamefont {Karabassov}, \citenamefont
  {Bobkova}, \citenamefont {Golubov},\ and\ \citenamefont
  {Vasenko}}]{Karabassov22}%
  \BibitemOpen
  \bibfield  {author} {\bibinfo {author} {\bibfnamefont {T.}~\bibnamefont
  {Karabassov}}, \bibinfo {author} {\bibfnamefont {I.~V.}\ \bibnamefont
  {Bobkova}}, \bibinfo {author} {\bibfnamefont {A.~A.}\ \bibnamefont
  {Golubov}}, \ and\ \bibinfo {author} {\bibfnamefont {A.~S.}\ \bibnamefont
  {Vasenko}},\ }\href {\doibase 10.1103/PhysRevB.106.224509} {\bibfield
  {journal} {\bibinfo  {journal} {Phys. Rev. B}\ }\textbf {\bibinfo {volume}
  {106}},\ \bibinfo {pages} {224509} (\bibinfo {year}
  {2022}{\natexlab{b}})}\BibitemShut {NoStop}%
\bibitem [{\citenamefont {Souto}\ \emph {et~al.}(2022)\citenamefont {Souto},
  \citenamefont {Leijnse},\ and\ \citenamefont {Schrade}}]{Souto22}%
  \BibitemOpen
  \bibfield  {author} {\bibinfo {author} {\bibfnamefont {R.~S.}\ \bibnamefont
  {Souto}}, \bibinfo {author} {\bibfnamefont {M.}~\bibnamefont {Leijnse}}, \
  and\ \bibinfo {author} {\bibfnamefont {C.}~\bibnamefont {Schrade}},\ }\href
  {\doibase 10.1103/PhysRevLett.129.267702} {\bibfield  {journal} {\bibinfo
  {journal} {Phys. Rev. Lett.}\ }\textbf {\bibinfo {volume} {129}},\ \bibinfo
  {pages} {267702} (\bibinfo {year} {2022})}\BibitemShut {NoStop}%
\bibitem [{\citenamefont {Zinkl}\ \emph {et~al.}(2022)\citenamefont {Zinkl},
  \citenamefont {Hamamoto},\ and\ \citenamefont {Sigrist}}]{Sigrist22}%
  \BibitemOpen
  \bibfield  {author} {\bibinfo {author} {\bibfnamefont {B.}~\bibnamefont
  {Zinkl}}, \bibinfo {author} {\bibfnamefont {K.}~\bibnamefont {Hamamoto}}, \
  and\ \bibinfo {author} {\bibfnamefont {M.}~\bibnamefont {Sigrist}},\ }\href
  {\doibase 10.1103/PhysRevResearch.4.033167} {\bibfield  {journal} {\bibinfo
  {journal} {Phys. Rev. Research}\ }\textbf {\bibinfo {volume} {4}},\ \bibinfo
  {pages} {033167} (\bibinfo {year} {2022})}\BibitemShut {NoStop}%
\bibitem [{\citenamefont {Fominov}\ and\ \citenamefont
  {Mikhailov}(2022)}]{Fominov22}%
  \BibitemOpen
  \bibfield  {author} {\bibinfo {author} {\bibfnamefont {Y.~V.}\ \bibnamefont
  {Fominov}}\ and\ \bibinfo {author} {\bibfnamefont {D.~S.}\ \bibnamefont
  {Mikhailov}},\ }\href {\doibase 10.1103/PhysRevB.106.134514} {\bibfield
  {journal} {\bibinfo  {journal} {Phys. Rev. B}\ }\textbf {\bibinfo {volume}
  {106}},\ \bibinfo {pages} {134514} (\bibinfo {year} {2022})}\BibitemShut
  {NoStop}%
\bibitem [{\citenamefont {Davydova}\ \emph {et~al.}(2022)\citenamefont
  {Davydova}, \citenamefont {Prembabu},\ and\ \citenamefont {Fu}}]{Davydova22}%
  \BibitemOpen
  \bibfield  {author} {\bibinfo {author} {\bibfnamefont {M.}~\bibnamefont
  {Davydova}}, \bibinfo {author} {\bibfnamefont {S.}~\bibnamefont {Prembabu}},
  \ and\ \bibinfo {author} {\bibfnamefont {L.}~\bibnamefont {Fu}},\ }\href
  {\doibase 10.1126/sciadv.abo0309} {\bibfield  {journal} {\bibinfo  {journal}
  {Science advances}\ }\textbf {\bibinfo {volume} {8}},\ \bibinfo {pages}
  {eabo0309} (\bibinfo {year} {2022})}\BibitemShut {NoStop}%
\bibitem [{\citenamefont {Tanaka}\ \emph {et~al.}(2022)\citenamefont {Tanaka},
  \citenamefont {Lu},\ and\ \citenamefont {Nagaosa}}]{Tanaka22}%
  \BibitemOpen
  \bibfield  {author} {\bibinfo {author} {\bibfnamefont {Y.}~\bibnamefont
  {Tanaka}}, \bibinfo {author} {\bibfnamefont {B.}~\bibnamefont {Lu}}, \ and\
  \bibinfo {author} {\bibfnamefont {N.}~\bibnamefont {Nagaosa}},\ }\href
  {\doibase 10.1103/PhysRevB.106.214524} {\bibfield  {journal} {\bibinfo
  {journal} {Phys. Rev. B}\ }\textbf {\bibinfo {volume} {106}},\ \bibinfo
  {pages} {214524} (\bibinfo {year} {2022})}\BibitemShut {NoStop}%
\bibitem [{\citenamefont {Wang}\ \emph {et~al.}(2022)\citenamefont {Wang},
  \citenamefont {Wang},\ and\ \citenamefont {Wu}}]{Wang2022}%
  \BibitemOpen
  \bibfield  {author} {\bibinfo {author} {\bibfnamefont {D.}~\bibnamefont
  {Wang}}, \bibinfo {author} {\bibfnamefont {Q.-H.}\ \bibnamefont {Wang}}, \
  and\ \bibinfo {author} {\bibfnamefont {C.}~\bibnamefont {Wu}},\ }\href
  {\doibase 10.48550/arXiv.2209.12646} {\enquote {\bibinfo {title} {Symmetry
  constraints on direct-current josephson diodes},}\ } (\bibinfo {year}
  {2022}),\ \Eprint {http://arxiv.org/abs/2209.12646} {arXiv:2209.12646}
  \BibitemShut {NoStop}%
\bibitem [{\citenamefont {Reynoso}\ \emph {et~al.}(2012)\citenamefont
  {Reynoso}, \citenamefont {Usaj}, \citenamefont {Balseiro}, \citenamefont
  {Feinberg},\ and\ \citenamefont {Avignon}}]{Reynoso12}%
  \BibitemOpen
  \bibfield  {author} {\bibinfo {author} {\bibfnamefont {A.~A.}\ \bibnamefont
  {Reynoso}}, \bibinfo {author} {\bibfnamefont {G.}~\bibnamefont {Usaj}},
  \bibinfo {author} {\bibfnamefont {C.~A.}\ \bibnamefont {Balseiro}}, \bibinfo
  {author} {\bibfnamefont {D.}~\bibnamefont {Feinberg}}, \ and\ \bibinfo
  {author} {\bibfnamefont {M.}~\bibnamefont {Avignon}},\ }\href {\doibase
  10.1103/PhysRevB.86.214519} {\bibfield  {journal} {\bibinfo  {journal} {Phys.
  Rev. B}\ }\textbf {\bibinfo {volume} {86}},\ \bibinfo {pages} {214519}
  (\bibinfo {year} {2012})}\BibitemShut {NoStop}%
\bibitem [{\citenamefont {Dolcini}\ \emph {et~al.}(2015)\citenamefont
  {Dolcini}, \citenamefont {Houzet},\ and\ \citenamefont {Meyer}}]{Meyer15}%
  \BibitemOpen
  \bibfield  {author} {\bibinfo {author} {\bibfnamefont {F.}~\bibnamefont
  {Dolcini}}, \bibinfo {author} {\bibfnamefont {M.}~\bibnamefont {Houzet}}, \
  and\ \bibinfo {author} {\bibfnamefont {J.~S.}\ \bibnamefont {Meyer}},\ }\href
  {\doibase 10.1103/PhysRevB.92.035428} {\bibfield  {journal} {\bibinfo
  {journal} {Phys. Rev. B}\ }\textbf {\bibinfo {volume} {92}},\ \bibinfo
  {pages} {035428} (\bibinfo {year} {2015})}\BibitemShut {NoStop}%
\bibitem [{\citenamefont {Alidoust}\ and\ \citenamefont
  {Hamzehpour}(2017)}]{Alidoust17}%
  \BibitemOpen
  \bibfield  {author} {\bibinfo {author} {\bibfnamefont {M.}~\bibnamefont
  {Alidoust}}\ and\ \bibinfo {author} {\bibfnamefont {H.}~\bibnamefont
  {Hamzehpour}},\ }\href {\doibase 10.1103/PhysRevB.96.165422} {\bibfield
  {journal} {\bibinfo  {journal} {Phys. Rev. B}\ }\textbf {\bibinfo {volume}
  {96}},\ \bibinfo {pages} {165422} (\bibinfo {year} {2017})}\BibitemShut
  {NoStop}%
\bibitem [{\citenamefont {Kulik}(1970)}]{Kulik}%
  \BibitemOpen
  \bibfield  {author} {\bibinfo {author} {\bibfnamefont {I.~O.}\ \bibnamefont
  {Kulik}},\ }\href@noop {} {\bibfield  {journal} {\bibinfo  {journal} {Sov.
  Phys. JETP}\ }\textbf {\bibinfo {volume} {30}},\ \bibinfo {pages} {944}
  (\bibinfo {year} {1970})}\BibitemShut {NoStop}%
\bibitem [{\citenamefont {Ishii}(1970)}]{Ishii70}%
  \BibitemOpen
  \bibfield  {author} {\bibinfo {author} {\bibfnamefont {C.}~\bibnamefont
  {Ishii}},\ }\href {\doibase 10.1143/PTP.44.1525} {\bibfield  {journal}
  {\bibinfo  {journal} {Prog. Theor. Phys.}\ }\textbf {\bibinfo {volume}
  {44}},\ \bibinfo {pages} {1525} (\bibinfo {year} {1970})}\BibitemShut
  {NoStop}%
\bibitem [{\citenamefont {Bardeen}\ and\ \citenamefont
  {Johnson}(1972)}]{Bardeen72}%
  \BibitemOpen
  \bibfield  {author} {\bibinfo {author} {\bibfnamefont {J.}~\bibnamefont
  {Bardeen}}\ and\ \bibinfo {author} {\bibfnamefont {J.~L.}\ \bibnamefont
  {Johnson}},\ }\href {\doibase 10.1103/PhysRevB.5.72} {\bibfield  {journal}
  {\bibinfo  {journal} {Phys. Rev. B}\ }\textbf {\bibinfo {volume} {5}},\
  \bibinfo {pages} {72} (\bibinfo {year} {1972})}\BibitemShut {NoStop}%
\bibitem [{\citenamefont {Fu}\ and\ \citenamefont {Kane}(2007)}]{Fu07}%
  \BibitemOpen
  \bibfield  {author} {\bibinfo {author} {\bibfnamefont {L.}~\bibnamefont
  {Fu}}\ and\ \bibinfo {author} {\bibfnamefont {C.~L.}\ \bibnamefont {Kane}},\
  }\href {\doibase 10.1103/PhysRevB.76.045302} {\bibfield  {journal} {\bibinfo
  {journal} {Phys. Rev. B}\ }\textbf {\bibinfo {volume} {76}},\ \bibinfo
  {pages} {045302} (\bibinfo {year} {2007})}\BibitemShut {NoStop}%
\bibitem [{\citenamefont {Zhang}\ \emph {et~al.}(2009)\citenamefont {Zhang},
  \citenamefont {Liu}, \citenamefont {Qi}, \citenamefont {Dai}, \citenamefont
  {Fang},\ and\ \citenamefont {Zhang}}]{Zhang09}%
  \BibitemOpen
  \bibfield  {author} {\bibinfo {author} {\bibfnamefont {H.}~\bibnamefont
  {Zhang}}, \bibinfo {author} {\bibfnamefont {C.-X.}\ \bibnamefont {Liu}},
  \bibinfo {author} {\bibfnamefont {X.-L.}\ \bibnamefont {Qi}}, \bibinfo
  {author} {\bibfnamefont {X.}~\bibnamefont {Dai}}, \bibinfo {author}
  {\bibfnamefont {Z.}~\bibnamefont {Fang}}, \ and\ \bibinfo {author}
  {\bibfnamefont {S.-C.}\ \bibnamefont {Zhang}},\ }\href {\doibase
  10.1038/nphys1270} {\bibfield  {journal} {\bibinfo  {journal} {Nature
  Physics}\ }\textbf {\bibinfo {volume} {5}},\ \bibinfo {pages} {438} (\bibinfo
  {year} {2009})}\BibitemShut {NoStop}%
\bibitem [{\citenamefont {Hsieh}\ \emph {et~al.}(2008)\citenamefont {Hsieh},
  \citenamefont {Qian}, \citenamefont {Wray}, \citenamefont {Xia},
  \citenamefont {Hor}, \citenamefont {Cava},\ and\ \citenamefont
  {Hasan}}]{Hsieh08}%
  \BibitemOpen
  \bibfield  {author} {\bibinfo {author} {\bibfnamefont {D.}~\bibnamefont
  {Hsieh}}, \bibinfo {author} {\bibfnamefont {D.}~\bibnamefont {Qian}},
  \bibinfo {author} {\bibfnamefont {L.}~\bibnamefont {Wray}}, \bibinfo {author}
  {\bibfnamefont {Y.}~\bibnamefont {Xia}}, \bibinfo {author} {\bibfnamefont
  {Y.}~\bibnamefont {Hor}}, \bibinfo {author} {\bibfnamefont {R.}~\bibnamefont
  {Cava}}, \ and\ \bibinfo {author} {\bibfnamefont {M.~Z.}\ \bibnamefont
  {Hasan}},\ }\href {\doibase 10.1038/nature06843} {\bibfield  {journal}
  {\bibinfo  {journal} {Nature}\ }\textbf {\bibinfo {volume} {452}},\ \bibinfo
  {pages} {970} (\bibinfo {year} {2008})}\BibitemShut {NoStop}%
\bibitem [{\citenamefont {Hsieh}\ \emph
  {et~al.}(2009{\natexlab{a}})\citenamefont {Hsieh}, \citenamefont {Xia},
  \citenamefont {Wray}, \citenamefont {Qian}, \citenamefont {Pal},
  \citenamefont {Dil}, \citenamefont {Osterwalder}, \citenamefont {Meier},
  \citenamefont {Bihlmayer}, \citenamefont {Kane}, \citenamefont {Hor},
  \citenamefont {Cava},\ and\ \citenamefont {Hasan}}]{Hsieh09}%
  \BibitemOpen
  \bibfield  {author} {\bibinfo {author} {\bibfnamefont {D.}~\bibnamefont
  {Hsieh}}, \bibinfo {author} {\bibfnamefont {Y.}~\bibnamefont {Xia}}, \bibinfo
  {author} {\bibfnamefont {L.}~\bibnamefont {Wray}}, \bibinfo {author}
  {\bibfnamefont {D.}~\bibnamefont {Qian}}, \bibinfo {author} {\bibfnamefont
  {A.}~\bibnamefont {Pal}}, \bibinfo {author} {\bibfnamefont {H.}~\bibnamefont
  {Dil}}, \bibinfo {author} {\bibfnamefont {J.}~\bibnamefont {Osterwalder}},
  \bibinfo {author} {\bibfnamefont {F.}~\bibnamefont {Meier}}, \bibinfo
  {author} {\bibfnamefont {G.}~\bibnamefont {Bihlmayer}}, \bibinfo {author}
  {\bibfnamefont {C.}~\bibnamefont {Kane}}, \bibinfo {author} {\bibfnamefont
  {Y.}~\bibnamefont {Hor}}, \bibinfo {author} {\bibfnamefont {R.}~\bibnamefont
  {Cava}}, \ and\ \bibinfo {author} {\bibfnamefont {M.~Z.}\ \bibnamefont
  {Hasan}},\ }\href {\doibase 10.1126/science.1167733} {\bibfield  {journal}
  {\bibinfo  {journal} {Science (New York, N.Y.)}\ }\textbf {\bibinfo {volume}
  {323}},\ \bibinfo {pages} {919} (\bibinfo {year}
  {2009}{\natexlab{a}})}\BibitemShut {NoStop}%
\bibitem [{\citenamefont {Xia}\ \emph {et~al.}(2009)\citenamefont {Xia},
  \citenamefont {Qian}, \citenamefont {Hsieh}, \citenamefont {Wray},
  \citenamefont {Pal}, \citenamefont {Lin}, \citenamefont {Bansil},
  \citenamefont {Grauer}, \citenamefont {Hor}, \citenamefont {Cava},\ and\
  \citenamefont {Hasan}}]{xia_2009}%
  \BibitemOpen
  \bibfield  {author} {\bibinfo {author} {\bibfnamefont {Y.}~\bibnamefont
  {Xia}}, \bibinfo {author} {\bibfnamefont {D.}~\bibnamefont {Qian}}, \bibinfo
  {author} {\bibfnamefont {D.}~\bibnamefont {Hsieh}}, \bibinfo {author}
  {\bibfnamefont {L.}~\bibnamefont {Wray}}, \bibinfo {author} {\bibfnamefont
  {A.}~\bibnamefont {Pal}}, \bibinfo {author} {\bibfnamefont {H.}~\bibnamefont
  {Lin}}, \bibinfo {author} {\bibfnamefont {A.}~\bibnamefont {Bansil}},
  \bibinfo {author} {\bibfnamefont {D.}~\bibnamefont {Grauer}}, \bibinfo
  {author} {\bibfnamefont {Y.~S.}\ \bibnamefont {Hor}}, \bibinfo {author}
  {\bibfnamefont {R.~J.}\ \bibnamefont {Cava}}, \ and\ \bibinfo {author}
  {\bibfnamefont {M.~Z.}\ \bibnamefont {Hasan}},\ }\href {\doibase
  10.1038/nphys1274} {\bibfield  {journal} {\bibinfo  {journal} {Nature
  Physics}\ }\textbf {\bibinfo {volume} {5}},\ \bibinfo {pages} {398} (\bibinfo
  {year} {2009})}\BibitemShut {NoStop}%
\bibitem [{\citenamefont {Chen}\ \emph {et~al.}(2009)\citenamefont {Chen},
  \citenamefont {Chu}, \citenamefont {Liu}, \citenamefont {Mo}, \citenamefont
  {Qi}, \citenamefont {Zhang}, \citenamefont {Lu}, \citenamefont {Dai},
  \citenamefont {Fang}, \citenamefont {Zhang}, \citenamefont {Fisher},
  \citenamefont {Hussain},\ and\ \citenamefont {Shen}}]{Chen09}%
  \BibitemOpen
  \bibfield  {author} {\bibinfo {author} {\bibfnamefont {Y.}~\bibnamefont
  {Chen}}, \bibinfo {author} {\bibfnamefont {J.-H.}\ \bibnamefont {Chu}},
  \bibinfo {author} {\bibfnamefont {Z.}~\bibnamefont {Liu}}, \bibinfo {author}
  {\bibfnamefont {S.-K.}\ \bibnamefont {Mo}}, \bibinfo {author} {\bibfnamefont
  {X.-L.}\ \bibnamefont {Qi}}, \bibinfo {author} {\bibfnamefont {H.-J.}\
  \bibnamefont {Zhang}}, \bibinfo {author} {\bibfnamefont {D.}~\bibnamefont
  {Lu}}, \bibinfo {author} {\bibfnamefont {X.}~\bibnamefont {Dai}}, \bibinfo
  {author} {\bibfnamefont {Z.}~\bibnamefont {Fang}}, \bibinfo {author}
  {\bibfnamefont {S.}~\bibnamefont {Zhang}}, \bibinfo {author} {\bibfnamefont
  {I.}~\bibnamefont {Fisher}}, \bibinfo {author} {\bibfnamefont
  {Z.}~\bibnamefont {Hussain}}, \ and\ \bibinfo {author} {\bibfnamefont
  {Z.-X.}\ \bibnamefont {Shen}},\ }\href {\doibase 10.1126/science.1173034}
  {\bibfield  {journal} {\bibinfo  {journal} {Science (New York, N.Y.)}\
  }\textbf {\bibinfo {volume} {325}},\ \bibinfo {pages} {178} (\bibinfo {year}
  {2009})}\BibitemShut {NoStop}%
\bibitem [{\citenamefont {Hsieh}\ \emph
  {et~al.}(2009{\natexlab{b}})\citenamefont {Hsieh}, \citenamefont {Xia},
  \citenamefont {Qian}, \citenamefont {Wray}, \citenamefont {Meier},
  \citenamefont {Dil}, \citenamefont {Osterwalder}, \citenamefont {Patthey},
  \citenamefont {Fedorov}, \citenamefont {Lin}, \citenamefont {Bansil},
  \citenamefont {Grauer}, \citenamefont {Hor}, \citenamefont {Cava},\ and\
  \citenamefont {Hasan}}]{Hasan09}%
  \BibitemOpen
  \bibfield  {author} {\bibinfo {author} {\bibfnamefont {D.}~\bibnamefont
  {Hsieh}}, \bibinfo {author} {\bibfnamefont {Y.}~\bibnamefont {Xia}}, \bibinfo
  {author} {\bibfnamefont {D.}~\bibnamefont {Qian}}, \bibinfo {author}
  {\bibfnamefont {L.}~\bibnamefont {Wray}}, \bibinfo {author} {\bibfnamefont
  {F.}~\bibnamefont {Meier}}, \bibinfo {author} {\bibfnamefont {J.~H.}\
  \bibnamefont {Dil}}, \bibinfo {author} {\bibfnamefont {J.}~\bibnamefont
  {Osterwalder}}, \bibinfo {author} {\bibfnamefont {L.}~\bibnamefont
  {Patthey}}, \bibinfo {author} {\bibfnamefont {A.~V.}\ \bibnamefont
  {Fedorov}}, \bibinfo {author} {\bibfnamefont {H.}~\bibnamefont {Lin}},
  \bibinfo {author} {\bibfnamefont {A.}~\bibnamefont {Bansil}}, \bibinfo
  {author} {\bibfnamefont {D.}~\bibnamefont {Grauer}}, \bibinfo {author}
  {\bibfnamefont {Y.~S.}\ \bibnamefont {Hor}}, \bibinfo {author} {\bibfnamefont
  {R.~J.}\ \bibnamefont {Cava}}, \ and\ \bibinfo {author} {\bibfnamefont
  {M.~Z.}\ \bibnamefont {Hasan}},\ }\href {\doibase
  10.1103/PhysRevLett.103.146401} {\bibfield  {journal} {\bibinfo  {journal}
  {Phys. Rev. Lett.}\ }\textbf {\bibinfo {volume} {103}},\ \bibinfo {pages}
  {146401} (\bibinfo {year} {2009}{\natexlab{b}})}\BibitemShut {NoStop}%
\bibitem [{\citenamefont {Ren}\ \emph {et~al.}(2010)\citenamefont {Ren},
  \citenamefont {Taskin}, \citenamefont {Sasaki}, \citenamefont {Segawa},\ and\
  \citenamefont {Ando}}]{Ren10}%
  \BibitemOpen
  \bibfield  {author} {\bibinfo {author} {\bibfnamefont {Z.}~\bibnamefont
  {Ren}}, \bibinfo {author} {\bibfnamefont {A.~A.}\ \bibnamefont {Taskin}},
  \bibinfo {author} {\bibfnamefont {S.}~\bibnamefont {Sasaki}}, \bibinfo
  {author} {\bibfnamefont {K.}~\bibnamefont {Segawa}}, \ and\ \bibinfo {author}
  {\bibfnamefont {Y.}~\bibnamefont {Ando}},\ }\href {\doibase
  10.1103/PhysRevB.82.241306} {\bibfield  {journal} {\bibinfo  {journal} {Phys.
  Rev. B}\ }\textbf {\bibinfo {volume} {82}},\ \bibinfo {pages} {241306}
  (\bibinfo {year} {2010})}\BibitemShut {NoStop}%
\bibitem [{\citenamefont {Tanaka}\ and\ \citenamefont
  {Kashiwaya}(1997)}]{Tanaka97}%
  \BibitemOpen
  \bibfield  {author} {\bibinfo {author} {\bibfnamefont {Y.}~\bibnamefont
  {Tanaka}}\ and\ \bibinfo {author} {\bibfnamefont {S.}~\bibnamefont
  {Kashiwaya}},\ }\href {\doibase 10.1103/PhysRevB.56.892} {\bibfield
  {journal} {\bibinfo  {journal} {Phys. Rev. B}\ }\textbf {\bibinfo {volume}
  {56}},\ \bibinfo {pages} {892} (\bibinfo {year} {1997})}\BibitemShut
  {NoStop}%
\bibitem [{\citenamefont {Burset}\ \emph {et~al.}(2015)\citenamefont {Burset},
  \citenamefont {Lu}, \citenamefont {Tkachov}, \citenamefont {Tanaka},
  \citenamefont {Hankiewicz},\ and\ \citenamefont {Trauzettel}}]{Pablo15}%
  \BibitemOpen
  \bibfield  {author} {\bibinfo {author} {\bibfnamefont {P.}~\bibnamefont
  {Burset}}, \bibinfo {author} {\bibfnamefont {B.}~\bibnamefont {Lu}}, \bibinfo
  {author} {\bibfnamefont {G.}~\bibnamefont {Tkachov}}, \bibinfo {author}
  {\bibfnamefont {Y.}~\bibnamefont {Tanaka}}, \bibinfo {author} {\bibfnamefont
  {E.~M.}\ \bibnamefont {Hankiewicz}}, \ and\ \bibinfo {author} {\bibfnamefont
  {B.}~\bibnamefont {Trauzettel}},\ }\href {\doibase
  10.1103/PhysRevB.92.205424} {\bibfield  {journal} {\bibinfo  {journal} {Phys.
  Rev. B}\ }\textbf {\bibinfo {volume} {92}},\ \bibinfo {pages} {205424}
  (\bibinfo {year} {2015})}\BibitemShut {NoStop}%
\bibitem [{\citenamefont {McMillan}(1968)}]{McMillan68}%
  \BibitemOpen
  \bibfield  {author} {\bibinfo {author} {\bibfnamefont {W.~L.}\ \bibnamefont
  {McMillan}},\ }\href {\doibase 10.1103/PhysRev.175.559} {\bibfield  {journal}
  {\bibinfo  {journal} {Phys. Rev.}\ }\textbf {\bibinfo {volume} {175}},\
  \bibinfo {pages} {559} (\bibinfo {year} {1968})}\BibitemShut {NoStop}%
\bibitem [{\citenamefont {Kashiwaya}\ and\ \citenamefont
  {Tanaka}(2000)}]{Kashiwaya_2000}%
  \BibitemOpen
  \bibfield  {author} {\bibinfo {author} {\bibfnamefont {S.}~\bibnamefont
  {Kashiwaya}}\ and\ \bibinfo {author} {\bibfnamefont {Y.}~\bibnamefont
  {Tanaka}},\ }\href {\doibase 10.1088/0034-4885/63/10/202} {\bibfield
  {journal} {\bibinfo  {journal} {Reports on Progress in Physics}\ }\textbf
  {\bibinfo {volume} {63}},\ \bibinfo {pages} {1641} (\bibinfo {year}
  {2000})}\BibitemShut {NoStop}%
\bibitem [{\citenamefont {Lu}\ \emph {et~al.}(2015{\natexlab{a}})\citenamefont
  {Lu}, \citenamefont {Burset}, \citenamefont {Yada},\ and\ \citenamefont
  {Tanaka}}]{Lu_2015}%
  \BibitemOpen
  \bibfield  {author} {\bibinfo {author} {\bibfnamefont {B.}~\bibnamefont
  {Lu}}, \bibinfo {author} {\bibfnamefont {P.}~\bibnamefont {Burset}}, \bibinfo
  {author} {\bibfnamefont {K.}~\bibnamefont {Yada}}, \ and\ \bibinfo {author}
  {\bibfnamefont {Y.}~\bibnamefont {Tanaka}},\ }\href {\doibase
  10.1088/0953-2048/28/10/105001} {\bibfield  {journal} {\bibinfo  {journal}
  {Superconductor Science and Technology}\ }\textbf {\bibinfo {volume} {28}},\
  \bibinfo {pages} {105001} (\bibinfo {year} {2015}{\natexlab{a}})}\BibitemShut
  {NoStop}%
\bibitem [{\citenamefont {Lu}\ \emph {et~al.}(2015{\natexlab{b}})\citenamefont
  {Lu}, \citenamefont {Yada}, \citenamefont {Golubov},\ and\ \citenamefont
  {Tanaka}}]{Yada15}%
  \BibitemOpen
  \bibfield  {author} {\bibinfo {author} {\bibfnamefont {B.}~\bibnamefont
  {Lu}}, \bibinfo {author} {\bibfnamefont {K.}~\bibnamefont {Yada}}, \bibinfo
  {author} {\bibfnamefont {A.~A.}\ \bibnamefont {Golubov}}, \ and\ \bibinfo
  {author} {\bibfnamefont {Y.}~\bibnamefont {Tanaka}},\ }\href {\doibase
  10.1103/PhysRevB.92.100503} {\bibfield  {journal} {\bibinfo  {journal} {Phys.
  Rev. B}\ }\textbf {\bibinfo {volume} {92}},\ \bibinfo {pages} {100503}
  (\bibinfo {year} {2015}{\natexlab{b}})}\BibitemShut {NoStop}%
\bibitem [{\citenamefont {Lu}\ and\ \citenamefont {Tanaka}(2018)}]{Lu18}%
  \BibitemOpen
  \bibfield  {author} {\bibinfo {author} {\bibfnamefont {B.}~\bibnamefont
  {Lu}}\ and\ \bibinfo {author} {\bibfnamefont {Y.}~\bibnamefont {Tanaka}},\
  }\href {\doibase 10.1098/rsta.2015.0246} {\bibfield  {journal} {\bibinfo
  {journal} {Philosophical Transactions of The Royal Society A Mathematical
  Physical and Engineering Sciences}\ }\textbf {\bibinfo {volume} {376}},\
  \bibinfo {pages} {20150246} (\bibinfo {year} {2018})}\BibitemShut {NoStop}%
\bibitem [{SM()}]{SM}%
  \BibitemOpen
  \href@noop {} {}\bibinfo {note} {See Supplemental Material (SM), including
  Refs.~\onlinecite{FT91,McMillan68,sun_17,Lee81,Ando91,Asano01}, where we
  derivate the supercurrent formula from Green's functions techniques, compute
  the Andreev bound states, and show a complementary tight-binding
  calculation.}\BibitemShut {Stop}%
\bibitem [{\citenamefont {Furusaki}\ and\ \citenamefont
  {Tsukada}(1991)}]{FT91}%
  \BibitemOpen
  \bibfield  {author} {\bibinfo {author} {\bibfnamefont {A.}~\bibnamefont
  {Furusaki}}\ and\ \bibinfo {author} {\bibfnamefont {M.}~\bibnamefont
  {Tsukada}},\ }\href {\doibase https://doi.org/10.1016/0038-1098(91)90201-6}
  {\bibfield  {journal} {\bibinfo  {journal} {Solid State Communications}\
  }\textbf {\bibinfo {volume} {78}},\ \bibinfo {pages} {299} (\bibinfo {year}
  {1991})}\BibitemShut {NoStop}%
\bibitem [{\citenamefont {Tanaka}\ \emph {et~al.}(2002)\citenamefont {Tanaka},
  \citenamefont {Tanuma}, \citenamefont {Kuroki},\ and\ \citenamefont
  {Kashiwaya}}]{Tanaka02}%
  \BibitemOpen
  \bibfield  {author} {\bibinfo {author} {\bibfnamefont {Y.}~\bibnamefont
  {Tanaka}}, \bibinfo {author} {\bibfnamefont {Y.}~\bibnamefont {Tanuma}},
  \bibinfo {author} {\bibfnamefont {K.}~\bibnamefont {Kuroki}}, \ and\ \bibinfo
  {author} {\bibfnamefont {S.}~\bibnamefont {Kashiwaya}},\ }\href {\doibase
  10.1143/JPSJ.71.2102} {\bibfield  {journal} {\bibinfo  {journal} {Journal of
  the Physical Society of Japan}\ }\textbf {\bibinfo {volume} {71}},\ \bibinfo
  {pages} {2102} (\bibinfo {year} {2002})}\BibitemShut {NoStop}%
\bibitem [{\citenamefont {Tkachov}\ \emph {et~al.}(2015)\citenamefont
  {Tkachov}, \citenamefont {Burset}, \citenamefont {Trauzettel},\ and\
  \citenamefont {Hankiewicz}}]{Tkachov15}%
  \BibitemOpen
  \bibfield  {author} {\bibinfo {author} {\bibfnamefont {G.}~\bibnamefont
  {Tkachov}}, \bibinfo {author} {\bibfnamefont {P.}~\bibnamefont {Burset}},
  \bibinfo {author} {\bibfnamefont {B.}~\bibnamefont {Trauzettel}}, \ and\
  \bibinfo {author} {\bibfnamefont {E.~M.}\ \bibnamefont {Hankiewicz}},\ }\href
  {\doibase 10.1103/PhysRevB.92.045408} {\bibfield  {journal} {\bibinfo
  {journal} {Phys. Rev. B}\ }\textbf {\bibinfo {volume} {92}},\ \bibinfo
  {pages} {045408} (\bibinfo {year} {2015})}\BibitemShut {NoStop}%
\bibitem [{\citenamefont {Tanaka}\ \emph {et~al.}(2009)\citenamefont {Tanaka},
  \citenamefont {Yokoyama},\ and\ \citenamefont {Nagaosa}}]{Tanaka09}%
  \BibitemOpen
  \bibfield  {author} {\bibinfo {author} {\bibfnamefont {Y.}~\bibnamefont
  {Tanaka}}, \bibinfo {author} {\bibfnamefont {T.}~\bibnamefont {Yokoyama}}, \
  and\ \bibinfo {author} {\bibfnamefont {N.}~\bibnamefont {Nagaosa}},\ }\href
  {\doibase 10.1103/PhysRevLett.103.107002} {\bibfield  {journal} {\bibinfo
  {journal} {Phys. Rev. Lett.}\ }\textbf {\bibinfo {volume} {103}},\ \bibinfo
  {pages} {107002} (\bibinfo {year} {2009})}\BibitemShut {NoStop}%
\bibitem [{\citenamefont {Xu}\ \emph {et~al.}(2014)\citenamefont {Xu},
  \citenamefont {Alidoust}, \citenamefont {Belopolski}, \citenamefont
  {Richardella}, \citenamefont {Liu}, \citenamefont {Neupane}, \citenamefont
  {Bian}, \citenamefont {Huang}, \citenamefont {Sankar}, \citenamefont {Fang},
  \citenamefont {Dellabetta}, \citenamefont {Dai}, \citenamefont {Li},
  \citenamefont {Gilbert}, \citenamefont {Chou}, \citenamefont {Samarth},\ and\
  \citenamefont {Hasan}}]{Xu2014}%
  \BibitemOpen
  \bibfield  {author} {\bibinfo {author} {\bibfnamefont {S.-Y.}\ \bibnamefont
  {Xu}}, \bibinfo {author} {\bibfnamefont {N.}~\bibnamefont {Alidoust}},
  \bibinfo {author} {\bibfnamefont {I.}~\bibnamefont {Belopolski}}, \bibinfo
  {author} {\bibfnamefont {A.}~\bibnamefont {Richardella}}, \bibinfo {author}
  {\bibfnamefont {C.}~\bibnamefont {Liu}}, \bibinfo {author} {\bibfnamefont
  {M.}~\bibnamefont {Neupane}}, \bibinfo {author} {\bibfnamefont
  {G.}~\bibnamefont {Bian}}, \bibinfo {author} {\bibfnamefont {S.-H.}\
  \bibnamefont {Huang}}, \bibinfo {author} {\bibfnamefont {R.}~\bibnamefont
  {Sankar}}, \bibinfo {author} {\bibfnamefont {C.}~\bibnamefont {Fang}},
  \bibinfo {author} {\bibfnamefont {B.}~\bibnamefont {Dellabetta}}, \bibinfo
  {author} {\bibfnamefont {W.}~\bibnamefont {Dai}}, \bibinfo {author}
  {\bibfnamefont {Q.}~\bibnamefont {Li}}, \bibinfo {author} {\bibfnamefont
  {M.~J.}\ \bibnamefont {Gilbert}}, \bibinfo {author} {\bibfnamefont
  {F.}~\bibnamefont {Chou}}, \bibinfo {author} {\bibfnamefont {N.}~\bibnamefont
  {Samarth}}, \ and\ \bibinfo {author} {\bibfnamefont {M.~Z.}\ \bibnamefont
  {Hasan}},\ }\href {\doibase 10.1038/nphys3139} {\bibfield  {journal}
  {\bibinfo  {journal} {Nature Physics}\ }\textbf {\bibinfo {volume} {10}},\
  \bibinfo {pages} {943} (\bibinfo {year} {2014})}\BibitemShut {NoStop}%
\bibitem [{\citenamefont {Zhu}\ \emph {et~al.}(2021)\citenamefont {Zhu},
  \citenamefont {Papaj}, \citenamefont {Nie}, \citenamefont {Xu}, \citenamefont
  {Gu}, \citenamefont {Yang}, \citenamefont {Guan}, \citenamefont {Wang},
  \citenamefont {Li}, \citenamefont {Liu}, \citenamefont {Luo}, \citenamefont
  {Zhu}, \citenamefont {Zheng}, \citenamefont {Fu},\ and\ \citenamefont
  {Jia}}]{JinFeng20}%
  \BibitemOpen
  \bibfield  {author} {\bibinfo {author} {\bibfnamefont {Z.}~\bibnamefont
  {Zhu}}, \bibinfo {author} {\bibfnamefont {M.}~\bibnamefont {Papaj}}, \bibinfo
  {author} {\bibfnamefont {X.-A.}\ \bibnamefont {Nie}}, \bibinfo {author}
  {\bibfnamefont {H.-K.}\ \bibnamefont {Xu}}, \bibinfo {author} {\bibfnamefont
  {Y.-S.}\ \bibnamefont {Gu}}, \bibinfo {author} {\bibfnamefont
  {X.}~\bibnamefont {Yang}}, \bibinfo {author} {\bibfnamefont {D.}~\bibnamefont
  {Guan}}, \bibinfo {author} {\bibfnamefont {S.}~\bibnamefont {Wang}}, \bibinfo
  {author} {\bibfnamefont {Y.}~\bibnamefont {Li}}, \bibinfo {author}
  {\bibfnamefont {C.}~\bibnamefont {Liu}}, \bibinfo {author} {\bibfnamefont
  {J.}~\bibnamefont {Luo}}, \bibinfo {author} {\bibfnamefont {a.}~\bibnamefont
  {Zhu}}, \bibinfo {author} {\bibfnamefont {H.}~\bibnamefont {Zheng}}, \bibinfo
  {author} {\bibfnamefont {L.}~\bibnamefont {Fu}}, \ and\ \bibinfo {author}
  {\bibfnamefont {J.-F.}\ \bibnamefont {Jia}},\ }\href {\doibase
  10.1126/science.abf1077} {\bibfield  {journal} {\bibinfo  {journal}
  {Science}\ }\textbf {\bibinfo {volume} {374}},\ \bibinfo {pages} {1381}
  (\bibinfo {year} {2021})}\BibitemShut {NoStop}%
\bibitem [{Note1()}]{Note1}%
  \BibitemOpen
  \bibinfo {note} {\protect \leavevmode {\protect \color {blue}Note that
  previous works, like, e.g., Refs.~\protect \rev@citealp {He_2022,Yuan22},
  propose that the diode effect is the result of Cooper pairs acquiring a
  finite momentum that, in principle, would be the same for all transport
  modes. In such a case, the sign of the quality factor does not change, even
  for long 2D junctions. The Doppler shift, by contrast, becomes
  angle-dependent thus allowing the angle-averaged quality factor to change
  sign}}\BibitemShut {NoStop}%
\bibitem [{\citenamefont {Tkachov}\ and\ \citenamefont
  {Hankiewicz}(2013)}]{Tkachov13}%
  \BibitemOpen
  \bibfield  {author} {\bibinfo {author} {\bibfnamefont {G.}~\bibnamefont
  {Tkachov}}\ and\ \bibinfo {author} {\bibfnamefont {E.~M.}\ \bibnamefont
  {Hankiewicz}},\ }\href {\doibase 10.1103/PhysRevB.88.075401} {\bibfield
  {journal} {\bibinfo  {journal} {Phys. Rev. B}\ }\textbf {\bibinfo {volume}
  {88}},\ \bibinfo {pages} {075401} (\bibinfo {year} {2013})}\BibitemShut
  {NoStop}%
\bibitem [{\citenamefont {Costa}\ \emph {et~al.}(2022)\citenamefont {Costa},
  \citenamefont {Baumgartner}, \citenamefont {Reinhardt}, \citenamefont
  {Berger}, \citenamefont {Gronin}, \citenamefont {Gardner}, \citenamefont
  {Lindemann}, \citenamefont {Manfra}, \citenamefont {Kochan}, \citenamefont
  {Fabian}, \citenamefont {Paradiso},\ and\ \citenamefont {Strunk}}]{Costa22}%
  \BibitemOpen
  \bibfield  {author} {\bibinfo {author} {\bibfnamefont {A.}~\bibnamefont
  {Costa}}, \bibinfo {author} {\bibfnamefont {C.}~\bibnamefont {Baumgartner}},
  \bibinfo {author} {\bibfnamefont {S.}~\bibnamefont {Reinhardt}}, \bibinfo
  {author} {\bibfnamefont {J.}~\bibnamefont {Berger}}, \bibinfo {author}
  {\bibfnamefont {S.}~\bibnamefont {Gronin}}, \bibinfo {author} {\bibfnamefont
  {G.}~\bibnamefont {Gardner}}, \bibinfo {author} {\bibfnamefont
  {T.}~\bibnamefont {Lindemann}}, \bibinfo {author} {\bibfnamefont
  {M.}~\bibnamefont {Manfra}}, \bibinfo {author} {\bibfnamefont
  {D.}~\bibnamefont {Kochan}}, \bibinfo {author} {\bibfnamefont
  {J.}~\bibnamefont {Fabian}}, \bibinfo {author} {\bibfnamefont
  {N.}~\bibnamefont {Paradiso}}, \ and\ \bibinfo {author} {\bibfnamefont
  {C.}~\bibnamefont {Strunk}},\ }\href@noop {} {\  (\bibinfo {year} {2022})},\
  \Eprint {http://arxiv.org/abs/2212.13460} {arXiv:2212.13460} \BibitemShut
  {NoStop}%
\bibitem [{\citenamefont {Zhou}\ \emph {et~al.}(2017)\citenamefont {Zhou},
  \citenamefont {Jiang}, \citenamefont {Xie},\ and\ \citenamefont
  {Sun}}]{sun_17}%
  \BibitemOpen
  \bibfield  {author} {\bibinfo {author} {\bibfnamefont {Y.-F.}\ \bibnamefont
  {Zhou}}, \bibinfo {author} {\bibfnamefont {H.}~\bibnamefont {Jiang}},
  \bibinfo {author} {\bibfnamefont {X.~C.}\ \bibnamefont {Xie}}, \ and\
  \bibinfo {author} {\bibfnamefont {Q.-F.}\ \bibnamefont {Sun}},\ }\href
  {\doibase 10.1103/PhysRevB.95.245137} {\bibfield  {journal} {\bibinfo
  {journal} {Phys. Rev. B}\ }\textbf {\bibinfo {volume} {95}},\ \bibinfo
  {pages} {245137} (\bibinfo {year} {2017})}\BibitemShut {NoStop}%
\bibitem [{\citenamefont {Lee}\ and\ \citenamefont {Fisher}(1981)}]{Lee81}%
  \BibitemOpen
  \bibfield  {author} {\bibinfo {author} {\bibfnamefont {P.~A.}\ \bibnamefont
  {Lee}}\ and\ \bibinfo {author} {\bibfnamefont {D.~S.}\ \bibnamefont
  {Fisher}},\ }\href {\doibase 10.1103/PhysRevLett.47.882} {\bibfield
  {journal} {\bibinfo  {journal} {Phys. Rev. Lett.}\ }\textbf {\bibinfo
  {volume} {47}},\ \bibinfo {pages} {882} (\bibinfo {year} {1981})}\BibitemShut
  {NoStop}%
\bibitem [{\citenamefont {Ando}(1991)}]{Ando91}%
  \BibitemOpen
  \bibfield  {author} {\bibinfo {author} {\bibfnamefont {T.}~\bibnamefont
  {Ando}},\ }\href {\doibase 10.1103/PhysRevB.44.8017} {\bibfield  {journal}
  {\bibinfo  {journal} {Phys. Rev. B}\ }\textbf {\bibinfo {volume} {44}},\
  \bibinfo {pages} {8017} (\bibinfo {year} {1991})}\BibitemShut {NoStop}%
\bibitem [{\citenamefont {Asano}(2001)}]{Asano01}%
  \BibitemOpen
  \bibfield  {author} {\bibinfo {author} {\bibfnamefont {Y.}~\bibnamefont
  {Asano}},\ }\href {\doibase 10.1103/PhysRevB.63.052512} {\bibfield  {journal}
  {\bibinfo  {journal} {Phys. Rev. B}\ }\textbf {\bibinfo {volume} {63}},\
  \bibinfo {pages} {052512} (\bibinfo {year} {2001})}\BibitemShut {NoStop}%
\end{thebibliography}%

\end{document}